\documentclass[manuscript]{aastex}
\usepackage{graphics}
\usepackage{epsfig}
\usepackage{amsbsy}

\slugcomment{to appear in ApJ}

\shorttitle{A NIR Fe\,{\sc ii} template for AGN}
\shortauthors{Garcia-Rissmann et al.}

\begin{document}


\title{A Near-Infrared Template Derived from I\,Zw\,1  \\
    for the Fe\,{\sc ii} Emission in Active Galaxies}


\author{A. Garcia-Rissmann\altaffilmark{} and A. Rodr\'{\i}guez-Ardila\altaffilmark{1}}
\affil{Laborat\'orio Nacional de Astrof\'{\i}sica,\\
 Rua Estados Unidos 154, 37504-364, Itajub\'a, MG, Brazil}

\author{T.A.A. Sigut\altaffilmark{1}}
\affil{The University of Western Ontario, \\
London, Ontario N6A 3K7 Canada}

\and

\author{A.K. Pradhan}
\affil{4055 McPherson Laboratory, The Ohio State University, \\
140 W. 18th Ave., Columbus, OH 43210-1173, USA}
\email{arissmann@lna.br}

\altaffiltext{1}{Visiting Astronomer at the Infrared Telescope Facility, which is operated
by the University of Hawaii under Cooperative Agreement no. NNX-08AE38A
with the National Aeronautics and Space Administration, Science Mission
Directorate, Planetary Astronomy Program.}


\begin{abstract}
In AGN spectra, a series of Fe\,{\sc ii} multiplets form a pseudo-continuum that extends
from the ultraviolet to the near-infrared (NIR). This emission is believed to originate in the Broad Line Region (BLR), and it has been known  
for a long time that pure photoionization fails to reproduce it  
in the most extreme cases, as does the collisional-excitation alone. The most recent models 
by \cite{SP03} include details of the Fe\,{\sc ii} ion microphysics and cover a wide
range in ionization parameter $\log U_{\rm ion}$= (-3.0~$\rightarrow$~-1.3) and density $\log
n_{\rm H}$ = (9.6~$\rightarrow$~12.6). With the aid of such models and a spectral
synthesis approach, we study for the first time in detail the NIR 
emission of I\,Zw\,1. The main goals are to confirm the role played by Ly$\alpha$ fluorescence
mechanisms in the production of the Fe\,{\sc ii} spectrum and to construct the
first semi-empirical NIR Fe\,{\sc ii} template that best represents this emission and can
be used to subtract it in other sources.  A good overall match between the
observed Fe\,{\sc ii}+Mg\,{\sc ii} features with those predicted by the best fitted model is
obtained, corroborating the Ly$\alpha$ fluorescence as a key
process to understand the Fe\,{\sc ii} spectrum. The best model is then adjusted by applying a 
deconvolution method on the observed Fe\,{\sc ii}+Mg\,{\sc ii} spectrum. The derived 
semi-empirical template is then fitted to the spectrum of Ark\,564, suitably 
reproducing its observed Fe\,{\sc ii}+Mg\,{\sc ii} emission. 
Our approach extends the current set of available Fe\,{\sc ii} templates into the NIR region.
\end{abstract}


\keywords{galaxies: active - galaxies: individual (I\,Zw\,1, Ark\,564) - 
infrared: general - quasars: emission lines - techniques: spectroscopic}



\section{Introduction}
\label{sec:Introduction}

The broad-line region (BLR) of active galactic nuclei (AGN) is thought 
to consist of a roughly spherical mist of cloudlets with characteristic 
densities in the range 10$^9$-10$^{12}$ cm$^{-3}$ and column densities 
of $\sim$10$^{24}$ cm$^{-2}$, surrouding a central source 
emitting ionizing radiation roughly isotropically \citep{sul00,gas09}. 
Despite the success of this traditional picture, in order to
explain the strengths of the BLR lines, the need for a high covering 
factor and the lack of Lyman continuum absorption, a BLR having
a flattened distribution at least for the low-ionization gas has been
proposed \citep{gas09}.

The BLR has been extensively studied 
from the X-rays to the NIR regime during the last two decades
\citep[see the reviews of][]{sul00,gas09}. One of the 
most puzzling aspects of the line spectra emitted by the BLR is the 
Fe\,{\sc ii} emission, whose numerous multiplets form a pseudo-continuum 
which extends from the UV to the optical region due to the blending of approximately
10$^5$ lines. This emission constitutes one of the most important contributors to the 
cooling of the BLR.

Indeed, the blending of several BLR emission lines, including the 
large number of Fe\,{\sc ii} emission multiplets, prevents a reliable study of
individual line profiles and the identification and measurement of weaker lines.
As blending is minimized in the class of objects known as Narrow-Line 
Seyfert~1 galaxies (NLS1s), their study can lead to a 
significantly more accurate study of the properties of the 
emission line region that is closer to the central source. 
\cite{BG92}, for instance, derived an optical template 
for the Fe\,{\sc ii} multiplets from I\,Zw\,1, which has served since then to 
adjust the Fe\,{\sc ii} strength of several other objects, after a proper scaling 
and convolution to match the BLR velocity dispersions (estimated from strong 
emission lines). The advent of HST UV quasar spectral data allowed
\cite{VW01} to extend the template method into 
the UV regime. They presented the first high S/N, high-resolution,
quasar empirical UV iron template spectrum ranging from rest frame 
1250~\AA\ to 3090~\AA, which is applicable to quasar data. The template 
was based on HST (archival) data of I\,Zw\,1.

The iron emission templates have importance not only
for our ability to measure and subtract the iron emission in quasar
spectra, but also as a tool through which we can study the iron
emission strength itself. Iron is a key coolant emitting
$\sim$25\% of the total energy output from the BLR \citep{WEA85}, emphasizing the
importance of including the iron emission in studies of the
BLR.

However, despite the wide use of such templates, most of the 
physical mechanisms that produce such lines remain under debate.
There have been a number of pioneering theoretical
investigations about the Fe\,{\sc ii} emission spectra in active galactic
nuclei (AGNs). For example, Phillips (1978a, 1978b) discussed
continuum pumping as one of the excitation mechanisms
that are responsible for that emission. \cite{NW83} and \cite{WEA85} calculated the strengths of 3407 Fe\,{\sc ii} emission
lines assuming collisional excitation and continuum fluorescence of Fe\,{\sc ii}, 
with radiative transfer in the spectral lines treated in the first-order escape 
probability approximation. They found a good fit to the overall
shape of Fe\,{\sc ii} features in the AGN UV and optical spectra but recognized
that the total strength of the Fe\,{\sc ii} emission is larger than the one
predicted by the models by a factor of $\sim$4. Other attemps made
by \cite{CS86} to solve the apparent weakness of the
Fe\,{\sc ii} emission using multi-component photoionization
models were also unsuccessful. These failures 
point out that not all the excitation mechanisms have been
taken into account or even that non-radiatively heated material with possibly 
even greater density exists within the BLR.

In order to solve the Fe\,{\sc ii} problem, \cite{P87} suggested Ly$\alpha$ 
flourescence as the main physical process involved in the production of 
Fe\,{\sc ii} lines. It takes advantage of the various near-coincidences between 
the wavelength of Ly$\alpha$ and the wavelengths corresponding to transitions 
between the levels of the $a^4D$ term and the $5p$ odd parity levels 
in Fe\,{\sc ii}, as described by \cite{JJ84}. The largest calculated 
transition probabilities from the odd $5p$ levels are those to $e^4D$ and 
$e^6D$, and cascades from these levels to odd parity levels at 5 eV and then to 
$a^6D$ and $a^4D$ would produce the bulk of  the Fe\,{\sc ii} spectrum located 
between 2000 \AA~ and 3000~\AA.

Model calculations including Ly$\alpha$ fluorescence as the excitation 
mechanism for the Fe\,{\sc ii} lines \citep{SP98,SP03} showed that this 
process is of fundamental importance in determining the 
strength of the Fe\,{\sc ii} emission. Previously, Fe\,{\sc ii} features in the intervals 
2400-2560 \AA~ and 2830-2900 \AA~ that originate from high excitation levels 
($\sim$10 eV) had been identified in the spectra of some AGNs \citep{GEA96,LEA97}, favouring the Ly$\alpha$ fluorescence scenario. The key 
feature to test this process, as predicted by \cite{SP98}, is 
significant Fe\,{\sc ii} emission in the wavelength range 8500-9500\AA, where the 
primary cascade lines from the upper $5p$ levels to $e^4D$ and $e^6D$ are 
located. Up to a few years ago, that emission had been elusive to observe in 
AGNs although they are common features in the spectra of some Galactic 
sources \citep{HP89,KEA94}.

Fortunately, sensitive NIR spectroscopy carried out on AGN samples 
during the last decade at moderate spectral resolution (R$\sim$800) 
revealed a wealth of emission lines from Fe\,{\sc ii} in a previously
unexplored wavelength region. For instance, \cite{RA02}, hereafter RA02, identified for the first time in four NLS1
galaxies (Ark\,564, \objectname[MRK 0335]{Mrk\,335}, \objectname[2MASX J19373299-0613046]{1H\,1934-063A} and \objectname[MRK 1044]{Mrk\,1044}) the 
strongest primary cascade lines of Ly$\alpha$ fluorescence predicted by 
\cite{SP98}. In addition, the secondary UV lines resulting
from the decay of the $e^4D$ and $e^6D$ levels are also present in these 
objects. Those results provided strong observational support to the 
hypothesis that Ly$\alpha$ fluorescence is, indeed, contributing to the 
emitted Fe\,{\sc ii} spectrum. Furthermore, RA02 reports the presence of 
the so-called 1$\mu$m Fe\,{\sc ii} lines (Fe\,{\sc ii} $\lambda$9997, $\lambda$10501, 
$\lambda$10862 and $\lambda$11126). These are the most prominent Fe\,{\sc ii} 
features observed in the rest wavelength interval 0.8-2.4 $\mu$m. 
The importance of that finding comes from the fact that nearly $\sim$50\% 
of the optical Fe\,{\sc ii} emission results from decays of the $z^4D^0$ and 
$z^4F^0$ levels, which are populated either by the transitions leading to the emission 
of the 1$\mu$m lines and the secondary UV lines mentioned above, or
by collisions from lower levels and direct photoionization. \cite{LEA08} reported similar 
findings in a sample of 23 well-known broad-emission line AGNs. 
They also confronted, for the first time, \cite{SP98} 
theoretical predictions of the NIR iron emission spectrum with 
observations. However, the prototypical I\,Zw\,1 was not included in their 
sample.

The only observation to date of I\,Zw\,1 in the NIR was reported by 
\cite{REA00}, hereafter RMPH. Their work clearly reveals Fe\,{\sc ii}\,
$\lambda$9997,\,$\lambda$10501,\,$\lambda$10862,\,$\lambda$11126. Based on the absence of
the crucial cascade lines that feed the common upper state
where the 1~$\mu$m Fe\,{\sc ii} lines originate (assuming Ly$\alpha$ fluorescence
as the dominant mechanism) as well as the relatively
low energy of that state, RMPH suggest that the observed
lines are collisionally excited. Note, however, that no individual 
identification of the Fe\,{\sc ii} lines in the 8500-9300~\AA\ interval has yet 
been done in I\,Zw\,1. 

For all said above, I\,Zw\,1 is a particularly good choice as
a test target to construct a semi-empirical NIR Fe\,{\sc ii}
template as it is so well studied, especially in terms of its
UV-optical iron emission. The strong and narrow Fe\,{\sc ii} emission
lines in I\,Zw\,1\footnote{Although I\,Zw\,1
is classified as a Seyfert~1 galaxy, its absolute luminosity 
($M_V$ = -23.8 for $H_0$ = 50 km~s$^{-1}$~Mpc$^{-1}$, $q_0$ = 0)
actually qualifies it as a low-luminosity quasar \citep{VCV91}.}, a conspicuous characteristic of NLS1 galaxies, make it 
an ideal object for the construction of
such a template and complements the ones previously published in
both the UV \citep{LEA97,VW01} and
in the optical region \citep{BG92,VCJV04}.

Primary cascading lines following Ly$\alpha$ fluorescence, 
previously confirmed by RA02 and \cite{LEA08} in other 
NLS1s, can also be studied and characterized in this source. 
The main 1~$\mu$m diagnostic lines have the advantage of not being 
heavily blended with other multiplets, as normally happens with the 
Fe\,{\sc ii} optical lines. The observation of such features in 
I~Zw~1 can serve as a useful benchmark for photoionization models, in 
particular, for models predicting the complex Fe\,{\sc ii} emission spectrum.  
Moreover, the proposed semi-empirical Fe\,{\sc ii} would allow the
subtraction of this emission in other AGNs. This is important for at 
least two reasons: to decontaminate other BLR features and to evaluate 
the amount of Fe\,{\sc ii} emission present in the NIR region, along with its
relationship to that of the UV and optical region.  

In this paper we describe the first detailed work to study the Fe\,{\sc ii} 
lines emitted by the BLR in I\,Zw\,1. The aim is 
twofold: {\it (i)} provide tight observational constraints to model the Fe\,{\sc ii} 
emission in this source and {\it (ii)} construct the first 
semi-empirical template in 
the NIR region. The structure of this paper is as follows: Section 2 describes 
the observations and data reduction. Section 3 discusses the characteristics of the theoretical models used in this work. In Section 4 we perform a template fitting to the observed spectrum of I\,Zw\,1, using the theoretical models described in the former section. Section 5 describes the 
construction of the semi-empirical Fe\,{\sc ii} template and tests it on the NIR spectrum of Ark\,564. A general discussion and conclusions are given in Sections 6 and 7, respectively. 

\section{Observations and Data Reduction}
\label{sec:observation}

Near-infrared spectra of I\,Zw\,1 were 
obtained at the NASA 3m Infrared Telescope Facility (IRTF) 
on the night of 23 October 2003. The SpeX 
spectrograph (Rayner et al. 2003) was used in the short cross-
dispersed mode (SXD, 0.8 - 2.4 $\mu$m). The detector employed
consisted of a 1024$\times$1024 ALADDIN 3 InSb array with a
spatial scale of 0.15$\arcsec$/pixel. A $0.8\arcsec\times15\arcsec$ 
slit oriented at the paralactic angle
to minimize differential refraction was used, providing a spectral 
resolution of 360 km\,s$^{-1}$. This value was determined both from
the arc lamp and the sky line spectra and was found to
be constant with wavelength along the observed spectra.

During the observations the seeing was $\sim$1$\arcsec$ in J. 
Observations were done nodding in an ABBA pattern with integration 
time of 120~s per frame and total on-source integration time 
of 28 minutes. After the galaxy, the A0V star SAO~92128 (V=7.38) 
was observed as telluric standard and flux calibrator. 
The spectral reduction, extraction and wavelength calibration 
procedures were performed using
SPEXTOOL \citep{CEA04}\footnote{SPEXTOOL is available from the
IRTF web site at http://irtf.ifa.hawaii.edu/Facility/spex/spex.html}, 
the in-house software developed and provided
by the SpeX team for the IRTF community. An aperture window 3$\arcsec$ wide was
employed to integrate all the signal from the galaxy nucleus along the
spatial direction. Extended emission is likely to be present
but it is outside the 3$\arcsec$ region. Indeed, the
FWHM of the I\,Zw\,1 light profile matches, within the natural seeing
fluctuations during the observations, that of the telluric
standard (0.91$\arcsec$ for the former and 0.89$\arcsec$ for
the latter in the $K$-band). The root-mean-square (RMS) of the dispersion
solution for the wavelength calibration was 0.17\AA.

The 1-D I\,Zw\,1 spectrum was then corrected for telluric absorption 
and flux calibrated using Xtellcor \citep{VCR03}, another in-house software developed by
the IRTF team. Finally, the different orders of the galaxy
spectrum were merged to form a single 1-D frame. It was
later corrected for the redshift of $z$=0.061105, determined 
from the average $z$ measured from the positions of Pa$\delta$, 
He\,{\sc i}~1.083$\mu$m, Pa$\beta$ and Br$\gamma$. A Galactic 
extinction correction of E(B-V)=0.065 \citep{SEA98} was 
applied.

\begin{figure}
\epsscale{.80}
\plotone{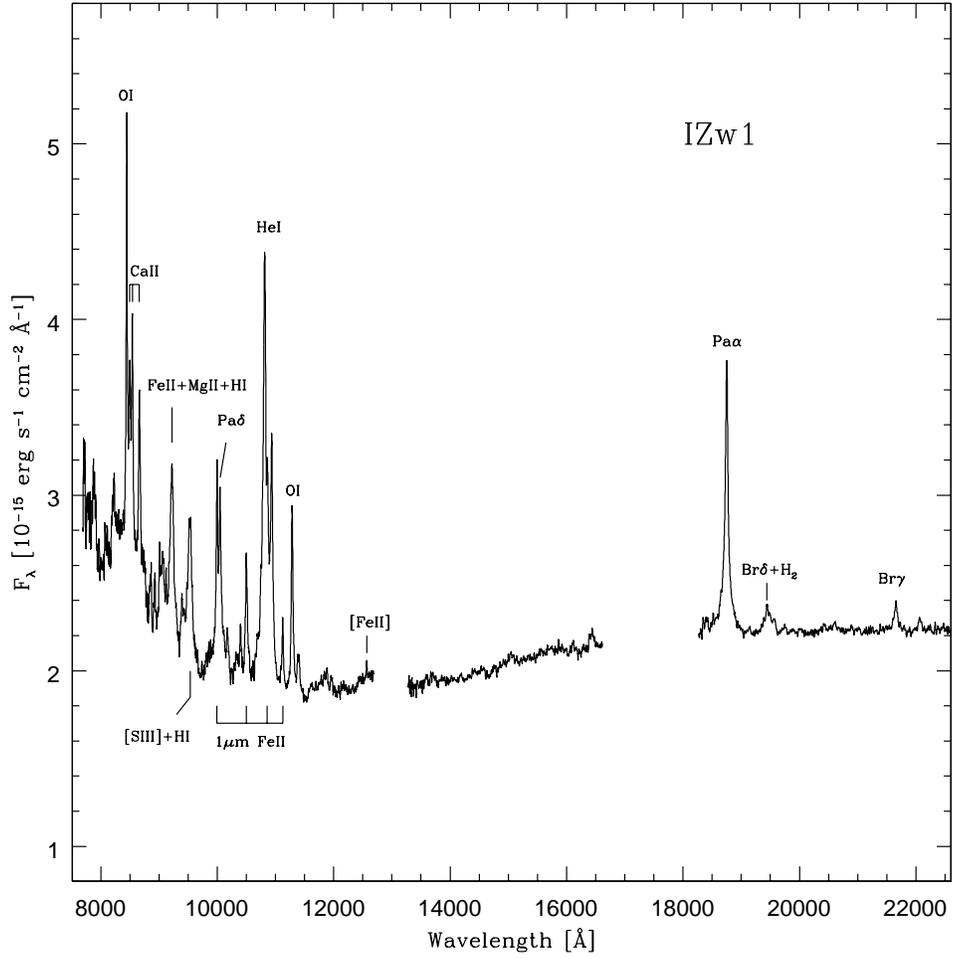}
\caption{NIR SpeX spectrum of I\,Zw\,1, from 8000~\AA\ to
22100\AA\ rest wavelength. Prominent emission
lines are identified.}.\label{izw1_irtf}
\end{figure}

Figure~\ref{izw1_irtf} shows the final 1D spectrum of I\,Zw\,1 
already calibrated by wavelength and flux. A visual comparison of 
our SpeX data with that of RMPH reveals an improvement in the 
spectral resolution along the 0.8-2.2 $\mu$m, allowing us to better
constrain most spectral features, in particular, those that
are heavily blended. Moreover, the higher S/N ($\geq$150) 
of our spectrum eases the identification of new emission 
features not detected before. It can also be seen that the
continuum flux scale is about 30\% lower in the $z, J$ and$ H$ band
and 50\% lower in the $K$ band than that of RMPH. This discrepancy
is very likely due to the differences in apertures between the
two observations, as our slit width is about 2.5$\times$ narrower, 
and therefore encompassing a much smaller contribution of the
host galaxy. The minimum in the continuum emission, 
at $\sim$13000~\AA\, is characteristic of AGNs and is interpreted
as due to a shift from a nonthermal continuum to the thermal dust 
emission that dominates at longer wavelengths 
(RMPH; RA02; Riffel, Rodr\'{\i}guez-Ardila \& Pastoriza 2006; 
 Landt et al. 2011). Note, however, that most line fluxes measured
in our observation (see Table~\ref{tab:el}) agree
within erros to that of RMPH.

Line identifications for the most conspicuous lines detected 
in the NIR spectrum of I\,Zw\,1 are indicated in Figure~\ref{izw1_irtf}.

It is easy to see from Figure~\ref{izw1_irtf}
that the NIR spectrum of I\,Zw\,1 is rich in
Fe\,{\sc ii} emission features. The so-called 1$\mu$m Fe\,{\sc ii}
lines, for instance, are particularly strong. Significant Fe\,{\sc ii} emission in the 8500-9500~\AA\
wavelength range, very likely produced by Ly$\alpha$ 
fluorescence as predicted by \cite{SP98}, is
also observed. Other prominent lines detected in I\,Zw\,1
include H\,{\sc i}, He\,{\sc i}, O\,{\sc i}
and Ca\,{\sc ii}. Moreover, forbidden lines of [Fe\,{\sc ii}], 
and [S\,{\sc iii}] and molecular H$_2$ were also detected.  

In the following sections we will discuss the method 
employed for the construction of the first semi-empirical 
NIR Fe\,{\sc ii} template published in the literature,
suitable to remove this emission in AGNs after a proper
subtraction of the continuum emission and scaling/
line broadening of the template. 

\section{The Fe\,{\sc ii} models}
\label{sec:models}

Theoretical models of Fe\,{\sc ii} including the NIR region are
rare in the literature. Up to our knowledge, all works
published before the year 1998 (Wills, Netzer \& Wills 1985 and references
therein) predicted emission line intensities 
for the UV and optical regions only. The main reason for 
ignoring the NIR is likely due to the fact that most model 
predictions pointed out to very weak 
Fe\,{\sc ii} emission in that region. Moreover, the lack of good S/N 
observations of AGNs redwards of 8000~\AA\ by that time prevented
a confrontation between models and observations.

\cite{SP98} proposed that the inclusion of Ly$\alpha$ fluorescence 
excitation process results in significant NIR Fe\,{\sc ii} emission in the 
region 8500-9500~\AA. In these calculations, a limited, non-LTE atomic model
with 262 fine structure levels, sufficiently large
for Ly$\alpha$ fluorescent excitation, was included. They
showed that Ly$\alpha$ excitation can be of fundamental
importance in enhancing the UV and optical Fe\,{\sc ii} fluxes.

Later, \cite{SP03} presented improved theoretical non-LTE
Fe\,{\sc ii} emission line strengths for physical conditions 
typical of active galactic nuclei with broad-line regions. In these
new set of models, updated to also include the Mg\,{\sc ii} ion (Sigut, 2004, private communication), the Fe\,{\sc ii} 
line strengths were computed with a precise 
treatment of radiative transfer using extensive and accurate atomic data 
from the Iron Project\footnote{NORAD database at www.astronomy.ohio-state.edu/$\sim$nahar} \citep{SNP04}. Excitation mechanisms for the Fe\,{\sc ii} emission 
included continuum fluorescence, collisional 
excitation, self-fluorescence among the Fe\,{\sc ii} transitions, and
fluorescent excitation by Ly$\alpha$ and Ly$\beta$. A larger Fe\,{\sc ii} 
atomic model consisting of 827 fine structure levels (including states to 
E$\sim$15 eV) was used to predict fluxes for approximately 23,000 Fe\,{\sc ii} 
transitions, covering most of the UV, optical, and NIR wavelengths of 
astrophysical interest. Detailed radiative transfer in the lines, including 
self-fluorescence overlap, was perfomed with an approximate $\Lambda$ operator 
scheme - see \cite{SP03} for details. 

Currently, owing to the complexity of the observed iron emission 
in AGNs, this emission is typically modeled using empirical templates derived 
from specific AGN spectra \citep{BG92,CB96}. A more recent example of application of this method consisted in deriving a Fe\,{\sc ii}-\,{\sc iii} template from high-quality UV spectra of the 
NLS1 galaxy I\,Zw\,1 \citep{VW01}. Such templates play a critical 
role in extracting a measure of the total iron emission from heavily 
blended and broadened AGN emission line spectra. Following this approach, and
taking advantage of the availability of the Fe\,{\sc ii} templates and  
the NIR spectrum of I\,Zw\,1, we will construct the first semi-empirical
Fe\,{\sc ii} template of that galaxy, complementing the ones
existing in the UV and optical regions. The approach that we will employ
consists of comparing the NIR spectrum of I\,Zw\,1 with a grid of
Fe\,{\sc ii}+Mg\,{\sc ii} models developed by Sigut \& Pradhan (2003), with these latter 
covering a wide range of ionization parameters ($U_{\rm ion}$=-1.3, -2 and -3 dex) and densities
($n_{\rm H}$ = 9.6, 10.6, 11.6 and 12.6 dex cm$^{-3}$). The internal 
cloud turbulent velocity, in all cases, was $V_{\rm tur}$=10~km\,s$^{-1}$.
Fe\,{\sc ii}+Mg\,{\sc ii} spectra were computed for BLR cloud models with 
typical conditions thought to exist in the Fe\,{\sc ii} emitting
clouds. The calculations have been made for traditional
clouds of a single specified density and ionization parameter,
as opposed to the more realistic locally optimally emitting
cloud models of \cite{BEA95}, as the main
interest is to study the interplay of the various iron emission excitation 
mechanisms and not the detailed structure of the BLR.

\begin{figure}
\epsscale{.80}
\plotone{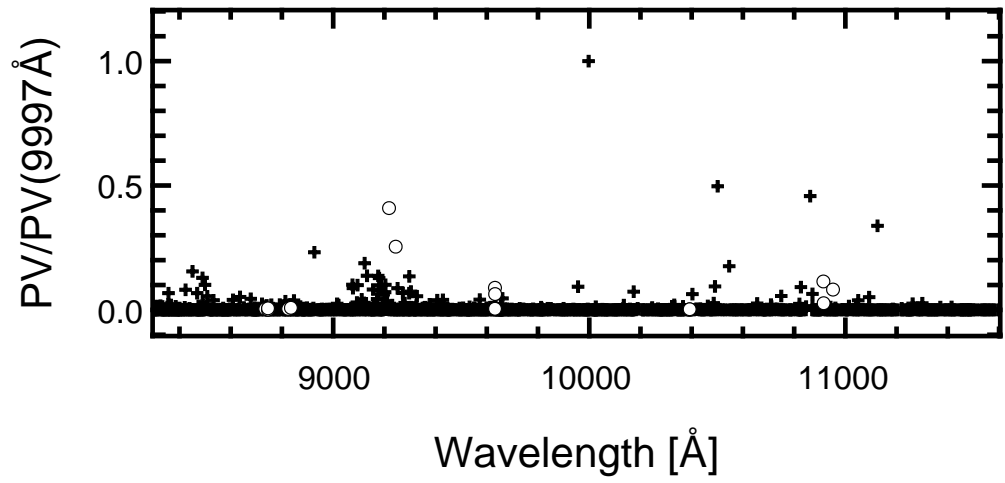}
\caption{Peak-to-valley variability of the Fe\,{\sc ii}+Mg\,{\sc ii} models at each wavelength, 
normalized by its value at 9997 \AA, for the 8300-11600 \AA~ region. Open circles refer to the Mg\,{\sc ii} emission.}.\label{p2v}
\end{figure}

Figure \ref{p2v} shows the peak-to-valley 
intensity variability among the Sigut \& Pradhan models 
versus wavelength. It shows a rather large variation in strength 
in some emission lines, 
such as those contained in the 8300-8500 and 9000-9400~\AA\ intervals 
(including a contribution from the Mg emission), Fe\,{\sc ii} at 8927, 
9997, 10502, 10862 and 11126 \AA, as well as Mg\,{\sc ii} at 9218 and 
9244 \AA. Indeed, the amplitude of the variation of these lines are 
about two orders of magnitude or more larger than the median peak-to-valley 
values.

\section{Analysis Procedure}
\label{sec:analysis}

In order to estimate a NIR template for I\,Zw\,1 a comparison of its 
observed emission line spectrum with that predicted by the available models 
was made. The best matching model can also indicate the most probable 
physical conditions of the BLR. As a by-product of the 
template fitting, we also determined the emission line positions/intensities 
of other BLR lines in order to minimize the residual RMS. The whole 
fitting procedure is decribed below.

The flux calibrated spectrum of I\,Zw\,1 was first continuum-subtracted 
in order to leave us with a pure emission line spectrum. For this purpose, 
a spline function was fitted to the continuum, choosing regions free of emission 
lines. The  $fit1d$ task of IRAF was used for this purpose, and chosen for simplicity. A more elaborate approach,
consisting of a simultaneous fitting of the intrinsic AGN continuum, stellar 
population template and dust is out from the scope of this paper.
Since the NIR spectrum is very rich in emission lines other than Fe\,{\sc ii}, 
it is also necessary to model them and then perform a multi-parametric fit.

The Paschen series was modeled using, as a reference, the observed 
Pa$\alpha$ (1.8751~$\mu$m) profile in velocity space.  In a first approach,
we set the scaling factors of individual lines as free 
parameters in the fit. Pa$\beta$ is located in a region of strong telluric 
contamination and, for that reason, was not considered. 
Preliminary tests produced results without physical significance (i.e. not justifiable 
by internal reddening and/or deviations of case B recombination rates). For that 
reason, we decided to set limits on the relative intensities of such lines consistent with decreasing 
values as moving towards the bluer part of the spectrum, starting from 
Pa$\gamma$. This behavior was forced through the adoption of reasonable boundary 
conditions in the their scaling parameters. Moreover,  Pa$_i$/Pa$_{i+1}$ line ratios 
in the 9000-12000 \AA~ range were allowed to have a 20-30\% error, given the 
uncertainties in subtracting the continuum. 

The permitted lines of Ca\,{\sc ii}\,$\lambda$8498,\,$\lambda$8542,\,$\lambda$8662,
He\,{\sc i}\,$\lambda$10830, He\,{\sc ii}\,$\lambda$10124 and 
O\,{\sc i}\,$\lambda$8446,\,$\lambda$11287 are also generally blended with the iron 
multiplets. As the bulk of all these lines is produced by the BLR, they 
were also modeled through the Pa$\alpha$ profile, allowing some broadening 
(by up to 80 km\,s$^{-1}$ using Gaussian filtering) to improve the results. 
The centroid position of He\,{\sc i} and He\,{\sc ii} were found to be 
blueshifted by -332 and -622~km\,s$^{-1}$, respectively. This may indicate
that the largest contribution of these two lines is produced in the NLR, in
agreement with the results of \cite{VCJV04}. They report
two regions emitting broad and blueshifted [O\,{\sc iii}] lines
in the optical region. One of the systems has V=-500~km\,s$^{-1}$, 
compatible to the line shift of the helium lines found here. The other system 
has V=-1450~km\,s$^{-1}$. This latter is not detected here,
very likely due to the lower spectral resolution of our data
and its relative weakness compared to the strength of the other system.  
The fact that the iron line profiles do not appear
asymmetric nor are significantly blueshifted indicates that
this emission does not originate in the outflowing gas itself.
This agrees with the results found by \cite{VW01}.

Forbidden 
lines present in this 
spectral region such as [Ca\,{\sc i}]\,$\lambda$9850 ($v_r \sim$213 km\,s$^{-1}$),  
the highly excited [S\,{\sc viii}]\,$\lambda$9913 ($v_r \sim$-785 km\,s$^{-1}$), 
[S\,{\sc iii}]\,$\lambda\lambda$9069,9532 (both with $v_r \sim$-300 km\,s$^{-1}$) 
and [S\,{\sc ii}]\,$\lambda$10280,\,$\lambda$10320 have all been assumed to 
have Lorentzian profiles with widths of 600 km\,s$^{-1}$. The blue shift found in most of these lines are 
compatible with the shift found by \cite{VCJV04} in the
optical spectrum, confirming the presence of a complex NLR in this object. We have fixed the FWHM\footnote{All FWHM listed in this 
work refer to the instrumental width, i.e. are not corrected from the intrinsic line width of 360 km\,s$^{-1}$.} 
of the forbidden lines because we verified that by progressively increasing their FWHM 
the residual RMS improved by clearly fitting through them also features of the Fe\,{\sc ii} models, 
a not desirable effect. Also, the [S\,{\sc iii}] doublet at
9069 and 9532 \AA~ were assumed to 
have their peak intensities constrained to the value of 2.4 
([S\,{\sc iii}]\,$\lambda$9532/[S\,{\sc iii}]\,$\lambda$9069), as determined by
atomic physics. The list of all modeled emission 
lines is shown in table \ref{tab:prof}. 

\begin{figure}
\epsscale{.80}
\plotone{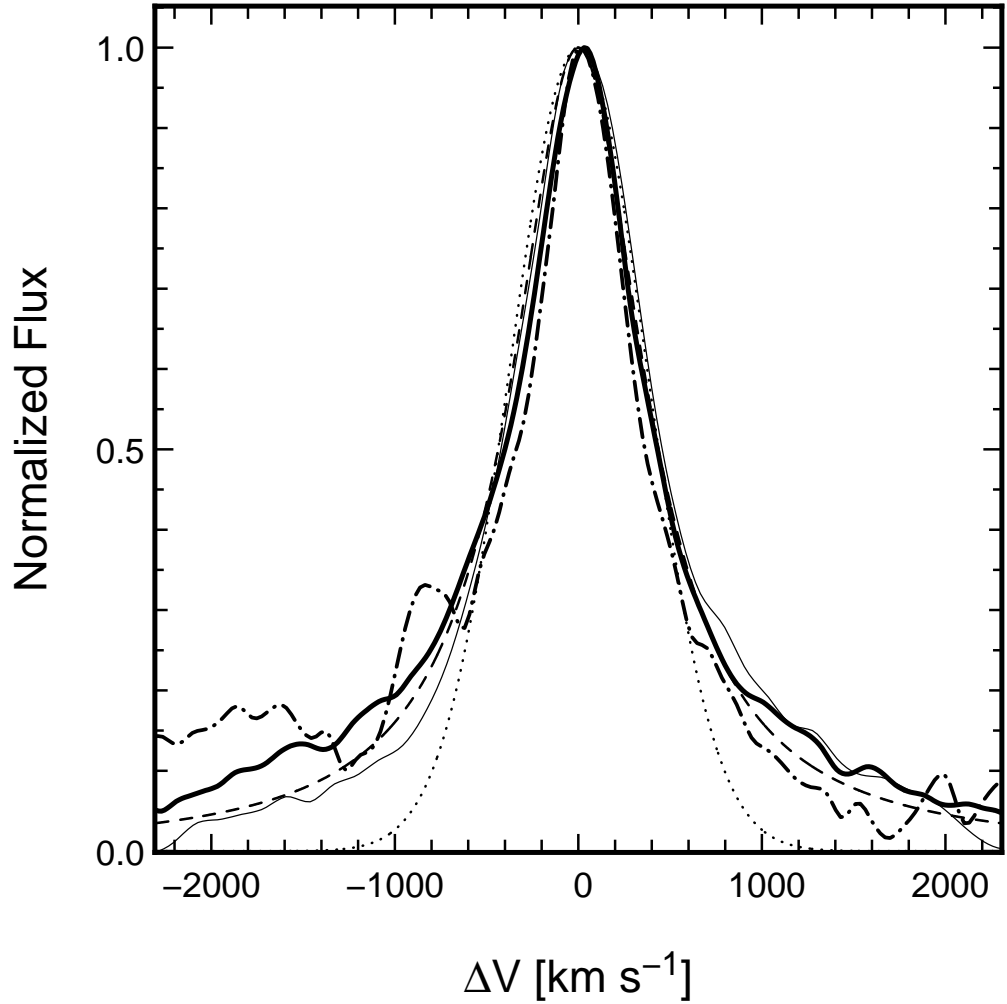}
\caption{The normalized O\,{\sc i}$\lambda$11286 \AA~(solid), Pa$\alpha$ (solid-bold) and Fe\,{\sc ii}$\lambda$11126 \AA~(dot-dashed) profiles. Gaussian (dotted) and Lorentzian (dashed) functions with the same FWHM of 
875 km\,s$^{-1}$ are also shown for comparison.}.\label{profOI}
\end{figure}

Our 12 available Fe\,{\sc ii}+Mg\,{\sc ii} (or simply the Fe\,{\sc ii}, 
when not considering the Mg\,{\sc ii} lines) 
theoretical templates had to be convolved with a line profile representative
of the BLR. For that purpose, it is useful to examine Figure \ref{profOI},
where the Fe\,{\sc ii}\,$\lambda$11126, O\,{\sc i}\,$\lambda$11287 and 
Pa$\alpha$ emission line profiles are plotted in the velocity space. 
It is easy to see that all three profiles can be well
represented by a Lorentzian function of FWHM = 875~km\,s$^{-1}$ (dashed curve). 
For comparison, the figure also shows a Gaussian profile (dotted) with the same 
FWHM as the Lorentzian. Clearly, the Gaussian fails at reproducing the extense wings 
observed in both lines. Because of the strong similarity of the BLR profiles 
with the Lorentzian curve, we adopted this theoretical profile to convolve 
our models. Notice that this system would be equivalent to the relatively broad 
L1 system of \cite{VCJV04}, associated to the BLR. 

Although it may sound appealing to use the form and width 
of Fe\,{\sc ii}\,$\lambda$9997 or ~$\lambda$10491+10502 to broaden the templates,
note that the former is heavily blended with Pa$\delta$ (narrow and broad components), 
[S\,{\sc viii}] and [C\,{\sc i}], and the latter is a blend of two lines very 
close in wavelength. Therefore, the characterization of these profiles
are more subject to uncertainties. As shown above, the form and width 
of either Fe\,{\sc ii}\,$\lambda$11126 or O\,{\sc i}\,$\lambda$11287
to broaden the Fe\,{\sc ii}+Mg\,{\sc ii} template provides a similar result.
This is consistent also with previous works  
\citep{RA02,MEA07} that presented consistent observational and theoretical
evidence, respectively, that both Fe\,{\sc ii} and O\,{\sc i} are originated in the 
same parcel of gas.

Let each line intensity of a given Fe\,{\sc ii}+Mg\,{\sc ii} 
(or only Fe\,{\sc ii}) template to be denoted 
by $l_{j}$ (a $\delta$-function at $\lambda_j$) and the Lorentzian velocity profile 
to be given by $P = P(v_{ij})$. The flux $f_i$ at a certain wavelenght $\lambda_i$ 
of the convolved template is then:
\begin{equation}
\label{eq:conv}
f_{i} \propto \sum_{all~ j} l_{j} \times P(v_{ij})
\end{equation}
with 
\begin{displaymath}
v_{ij} = c~ [ (\lambda_i - \lambda_j)/\lambda_j ]
\end{displaymath}
\noindent where $c$ is the speed of light. The proportionality constant on equation 
\ref{eq:conv} is a parameter of the fit. 

\begin{figure}
\epsscale{.80}
\plotone{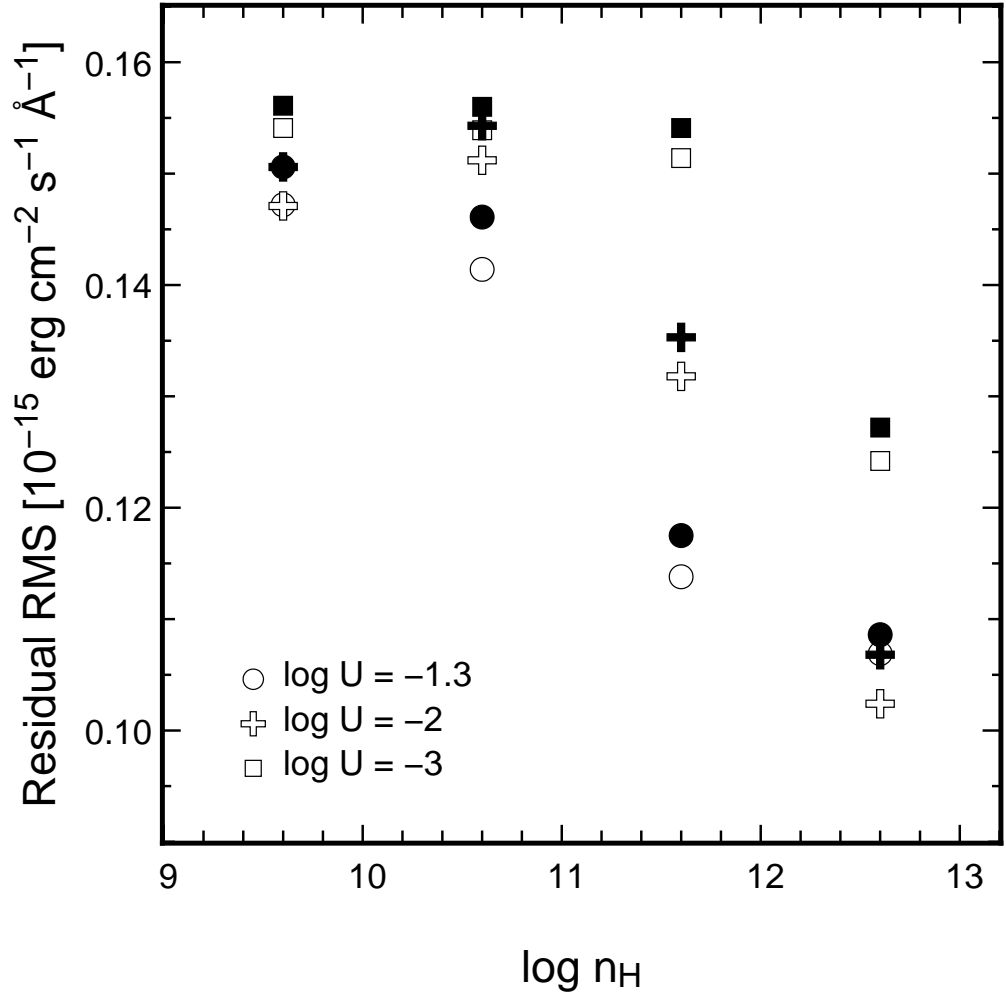}
\caption{RMS of the residual spectrum of I\,Zw\,1 after subtraction of fitted 
Fe\,{\sc ii}+Mg\,{\sc ii} models and emission lines, in the region between 8300 and 11600 \AA. 
Solid symbols denote the results obtained using only the Fe\,{\sc ii} multiplets 
(Mg\,{\sc ii} neglected).}.\label{rms}
\end{figure}

Once defined all the emission line profiles, we proceeded with the 
least-squares fit of all scaling parameters (19 in total for each model,
correponding to the lines listed in Table~\ref{tab:prof} plus the convolved template scaling factor)\footnote{Notice that the S\,{\sc iii} lines count as one, since they are constrained in intensity ratio.}. 
This was carried out using the Fortran routines of the Minuit v. 94.1 software, 
available at the CERN library\footnote{wwwasdoc.web.cern.ch/wwwasdoc/minuit/minmain.html}. 
The boundary conditions provided H\,{\sc i} line ratios (Pa$\gamma$/Pa$\delta$, 
Pa$\delta$/Pa8, Pa8/Pa9, Pa9/Pa10) with an average of $\sim$1.7, 
consistent with the theoretical Paschen decrements within 20$-$30\% error. 
Note that the reduced-$\chi^2$ provided by the routine does not take into 
account possible sources of errors such as residuals from telluric absorption 
corrections and continuum subtraction. For this reason, we
used the RMS of the residuals as a measure for the quality of the fit. 
The residual RMS obtained from the fit of each of the 12 models, calculated in the 
the main region of interest (8300$-$11600~\AA), is shown in Figure~\ref{rms}. These
results are used as a discriminant of the best suitable 
 model.
 
Models with medium/low ionization parameters and large densities such 
as (log~$U_{\rm ion}$=-2.0, log~$n_{\rm H}$=12.6) and (log~$U_{\rm ion}$=-1.3, log~$n_{\rm H}$=12.6) 
are the most successful ones, having residual RMS between 15-20\% smaller with respect 
to the average (around 30\% smaller considering the worst models). 
For consistency, we decided to check the effect of the Mg\,{\sc ii} emission 
in the fits. For this purpose, we performed the same fitting procedure 
using only the Fe\,{\sc ii} multiplets in the models. This approach is 
justified by the large uncertainties concerning the Mg\,{\sc ii} strength 
in the models, specially in the spectral region of 9200~\AA. The excitation 
mechanism for these Mg\,{\sc ii} lines is Ly$\beta$ fluorescence, and the 
uncertainties come from the small transition probabilities between the levels 
3s $^2$S $\rightarrow$ 5p $^2$P$^o$. Moreover, the pumping source function for Ly$\beta$ is more 
uncertain than for Ly$\alpha$ due to the fact that photons in Ly$\beta$ 
can also cycle through H$\alpha$, and this ``cross-redistribution'' can be important. 

Considering that the number of Mg\,{\sc ii} lines included in the 
Fe\,{\sc ii}+Mg\,{\sc ii} models is less than 1\% of that of Fe\,{\sc ii},
the exclusion of Mg\,{\sc ii} does not affect significantly the fit, as expected. The RMS 
values of the fit residuals considering only the Fe\,{\sc ii} multiplets 
are also shown in Figure~\ref{rms} as solid symbols. The results are very 
similar to those obtained using the composite Fe\,{\sc ii}+Mg\,{\sc ii} models, 
favoring a high-density and moderate/low ionization parameter.

Another consistency check in the results was made considering only the 8300-11600~\AA~ 
region in the fit, where the main Fe\,{\sc ii} diagnostic lines are located. 
In the above interval only Pa$\gamma$ to Pa10 were 
included in the fit, reducing the number of free parameters to 18. It turned 
out that either by fitting the larger 8300-20000~\AA\ or just the 
8300-11600~\AA\ interval one obtains similar results. For this reason, and 
also because of the lack of significant Fe\,{\sc ii}+Mg\,{\sc ii} 
emission redwards of 11600 \AA, we will concentrate in the remainder of 
this paper to this smaller spectral region.

Figures \ref{pfit1} and \ref{pfit2} show the best fits for all the models. 
Table \ref{tab:el} lists the emission line intensities derived from such fits.

\section{Semi-empirical NIR Fe\,{\sc ii}+Mg\,{\sc ii} Template}

In the previous section we found that the model with log~$U$=-2 and 
log~$n_{\rm H}$=12.6 is the one that best represents the observed Fe\,{\sc ii}+Mg\,{\sc ii} emission in 
I\,Zw\,1. Although not crucial at this point, it is important to question if that 
model is consistent with the physical conditions of a Fe\,{\sc ii}
emission region believed to exist in AGNs. 

\cite{jol91} computed purely collisional models
showing that low temperature ($T<$8000~K), high density
($n_{\rm H}>10^{11}~$cm$^{-3}$) and high column density ($N$(H)$>10^{22}$~cm$^{-2}$)
clouds provide Fe\,{\sc ii}$_{\rm \,opt}$/H$\beta$ in good agreement with
observations of Seyfert~1 galaxies. Detailed modeling of the Fe\,{\sc ii} using 
Cloudy and including Ly$\alpha$ 
fluorescence as one of the excitation mechanisms made by \cite{BEA04} 
points out to similar conditions. Indeed, densities between 9~$<~n_{\rm H}~<~13$ 
and log~$U_{\rm ion}$=-1.4 are consistent with observations. Additional observational
evidence that strong iron (Fe\,{\sc ii} and Fe\,{\sc iii}) emission may be
connected with high densities was provided by \citet{HB86,jol91,bal96,law97}
and \citet{kur00}. From that we can see that the physical conditions of the
best matching model for I\,Zw\,1 are, in general terms, representative of the Fe\,{\sc ii}
emitting region in AGNs. 

In the remainder of this section we analyze in detail the best 
Fe\,{\sc ii}+Mg\,{\sc ii} template and propose modifications to it, based on 
the mismatches between the observed spectrum and 
the convolved model. We then present a semi-empirical template capable of minimizing the residual 
RMS in the NIR window of 8300-11600 \AA~ for I\,Zw\,1.

The equation \ref{eq:conv} for the convolution of a given model can be succinctly 
rewritten in a matricial form:
\begin{equation}
\label{eq:conv2}
\boldsymbol{\mathit{f}} = \mbox{\bf P}~\boldsymbol{\ell}
\end{equation}  
\noindent 

where $\boldsymbol{\mathit{f}}$ denotes a spectrum, the N-length vector 
of  Fe\,{\sc ii}+Mg\,{\sc ii} fluxes for a given wavelength range, 
$\mbox{\bf P}$ the N$\times$M convolution matrix and $\boldsymbol{\ell}$ 
the M-length vector with the set of Fe\,{\sc ii} and Mg\,{\sc ii} line 
intensities which contribute to the wavelength range comprised by 
$\boldsymbol{\mathit{f}}$. Neglecting any measurement errors, we can estimate 
$\boldsymbol{\mathit{f}}$ from the 
residuals of the observed spectrum after the subtraction of all emission 
lines (except Fe\,{\sc ii} and Mg\,{\sc ii}), which we call hereafter $\boldsymbol{\mathit{f}}_{obs}$. 
The convolution matrix can be easily built from the Lorentzian velocity 
profile (FWHM=875 km s$^ {-1}$) and the $v_{ij}$ differences between each 
given multiplet line $l_j$ and the wavelength which corresponds to $f_i$. 
Adjusting the template line intensities to match as close as possible the 
observed residuals then constitutes a deconvolution problem, for which 
several solving approaches exist (see, for instance, a discussion on 
deconvolution techniques in Thi\`{e}baut 2005).  

The number of Fe\,{\sc ii}+Mg\,{\sc ii} lines comprised between 8300 and 
11600~\AA~ are 1529, out of which only a few contribute significantly to 
the emission spectrum. The Wiener deconvolution was chosen for our problem for 
its simplicity. However, it can lead to an unrealistic overdensity of line 
contributions, along with some unphysical results, if we consider all of the 
1529 line intensities as free parameters. For this reason, we have decided 
to build a $\boldsymbol{\ell}^*$ vector containing only the contribution 
of multiplets with the $\approx$3\% largest intensity contributions 
(selected from the best fitted model), plus some others whose intensities are
clearly underestimated in the models (e.g. the unusual two-peaked bump around 
11400 \AA; see discussion below). In practice, this means zeroing the 
elements in the $\boldsymbol{\ell}$ vector (to be multiplied by $\mbox{\bf P}$) which fall outside the selection 
criteria. The intensity threshold for building $\boldsymbol{\ell}^*$ lies at 
about 4\% of the peak intensity in this wavelength range, and it is at least twice higher 
than the template median intensity. This criterium produces a $\boldsymbol{\ell}^*$ with 51 elements.

In order to account for the spectral contribution of the 
Fe\,{\sc ii}+Mg\,{\sc ii} multiplets not present in the $\boldsymbol{\ell}^*$ 
vector, which we call hereafter the ``background'' spectrum $\boldsymbol{\mathit{f}}_{b}$, 
we first subtract their emission\footnote{The ``background'' spectrum $\boldsymbol{\mathit{f}}_{b}$ is obtained by convolving the low intensity lines selected from the best 
model with the Lorentzian profile.} from the input spectrum. 
This allows us to rewrite equation (\ref{eq:conv2}) as:
\begin{equation}
\boldsymbol{\mathit{f}}_{obs}^* = \boldsymbol{\mathit{f}}_{obs} - \boldsymbol{\mathit{f}}_{b} = \mbox{\bf P}~\boldsymbol{\ell}^*
\end{equation}
Note that $\boldsymbol{\mathit{f}}_{b}$ is scaled accordingly to the 
fitted parameter of Fe\,{\sc ii}+Mg\,{\sc ii} found in the previous section. 
The line intensities in the vector $\boldsymbol{\ell}^*$ found from 
the deconvolution of $\boldsymbol{\mathit{f}}_{obs}^*$ plus the ``background'' 
subset of line intensities provide what we call the semi-empirical 
Fe\,{\sc ii}+Mg\,{\sc ii} template.

\begin{figure}
\epsscale{.80}
\plotone{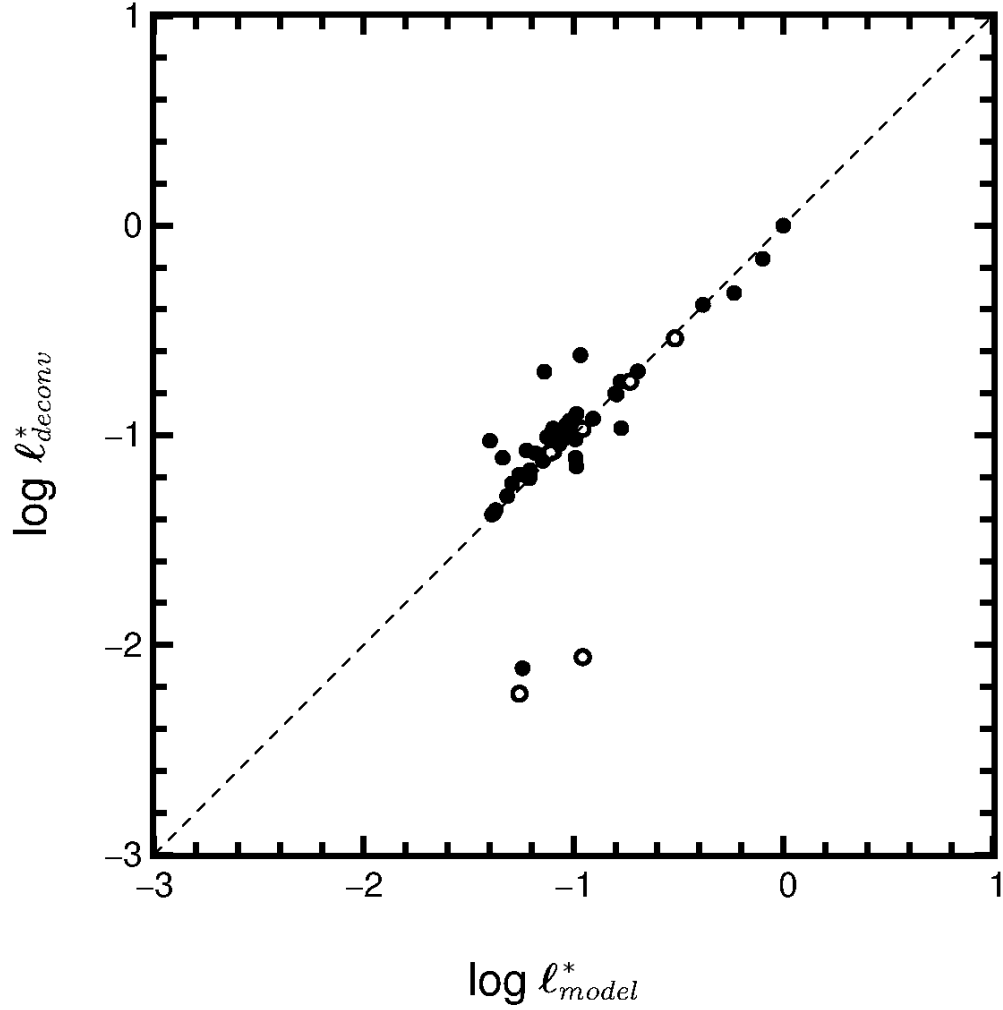}
\caption{Comparison between the 51 elements of the input intensity vector $\boldsymbol{\ell}^*_{model}$ (=model) with those of the output one $\boldsymbol{\ell}^*_{deconv}$ (=deconvolved), normalized by the intensities at 9997\AA. Filled and open circles denote Fe\,{\sc ii} and Mg\,{\sc ii} lines, respectively. The Wiener deconvolution gives a consistent result, specially for the strongest contributions. The dashed line represents the locus of the 1:1 relation.}\label{fig:dec}
\end{figure}

A quick test used as $\boldsymbol{\mathit{f}}_{obs}$ a spectrum produced by convolving our best model, which we called $\boldsymbol{\mathit{f}}_{model}$. The deconvolution technique described above was then applied to check whether we could reliably recover the original $\boldsymbol{\ell}^*$, and consequently $\boldsymbol{\ell}$. Convolving this ``deconvolved'' vector then produced $\boldsymbol{\mathit{f}}_{deconv}$. The difference of this latter with  $\boldsymbol{\mathit{f}}_{model}$ showed a RMS of less than 2\% (relative to the intensity at 9997 \AA). The reliability of this result is also shown in 
Figure~\ref{fig:dec}, where the initial normalized model intensities 
($\boldsymbol{\ell}^*_{model}$) are plotted against the ones found from the 
deconvolution ($\boldsymbol{\ell}^*_{deconv}$). Despite of some scattering, 
specially for the less intense lines, the agreement is very good.

When tackling real data, we decided to make a small tuning to the $\boldsymbol{\ell}^*$ vector. In order to 
conserve the physical interpretation given by the 
best model (-2.0,12.6) with predictions about the intensities of the most prominent Fe\,{\sc ii} lines, 
we decided to discard them from the $\boldsymbol{\ell}^*$ vector. 
In practice this means that Fe\,{\sc ii}\,$\lambda$9997,\,$\lambda$10501,\,$\lambda$10862, 
\,$\lambda$11126 (and the close neighbors Fe\,{\sc ii}\,$\lambda$10491,\,$\lambda$10871) 
had their original values preserved, and have been included in the $\boldsymbol{\mathit{f}}_{b}$ contribution.
We should also make a brief remark on the 
relatively strong two-peaked bump of Fe\,{\sc ii} observed at $\sim$11400~\AA: this feature is underpredicted by any of the models, 
and because of its strength, we ruled out the possibility of being 
part of the spectrum noise or a feature introduced by the 
O\,{\sc i}\,$\lambda$11287 profile fitting. Telluric absorption 
corrections are not critical in this region and, besides that, in 
the model templates one can notice two small peaks around this region, 
not exactly coincident in wavelength with the observed ones. These 
arguments give us support to believe that these features, located around  
11381 and 11402 \AA, are real. Notice that these wavelengths, 
estimated by visual inspection, are close to 11380.32 and 11403.54\AA, 
which refer to features that appear in absorption in the models. 
We decided to include these two hypothetical lines in the 
$\boldsymbol{\ell}^*$ vector, as well as those around 10686~\AA, 
clearly underestimated in the models but visible in 
$\boldsymbol{\mathit{f}}_{obs}^*$.  This left us with a final 50-element $\boldsymbol{\ell}^*$ vector.

We then applied the deconvolution to the observed spectrum 
(after the $\boldsymbol{\mathit{f}}_{b}$ subtraction). 
The intensities obtained from the deconvolution method applied to 
the observed Fe\,{\sc ii}+Mg\,{\sc ii} I\,Zw\,1 spectrum are shown 
in Table \ref{tab:dec} together with those given by the best  
model. These values are all normalized with respect to the intensity 
of the Fe\,{\sc ii}\,$\lambda$9997 line. For building the semi-empirical 
Fe\,{\sc ii}+Mg\,{\sc ii} spectrum we convolved this newly derived 
intensity vector with the $\mbox{\bf P}$ matrix and added back 
the background contribution.

\begin{figure}
\epsscale{.80}
\plotone{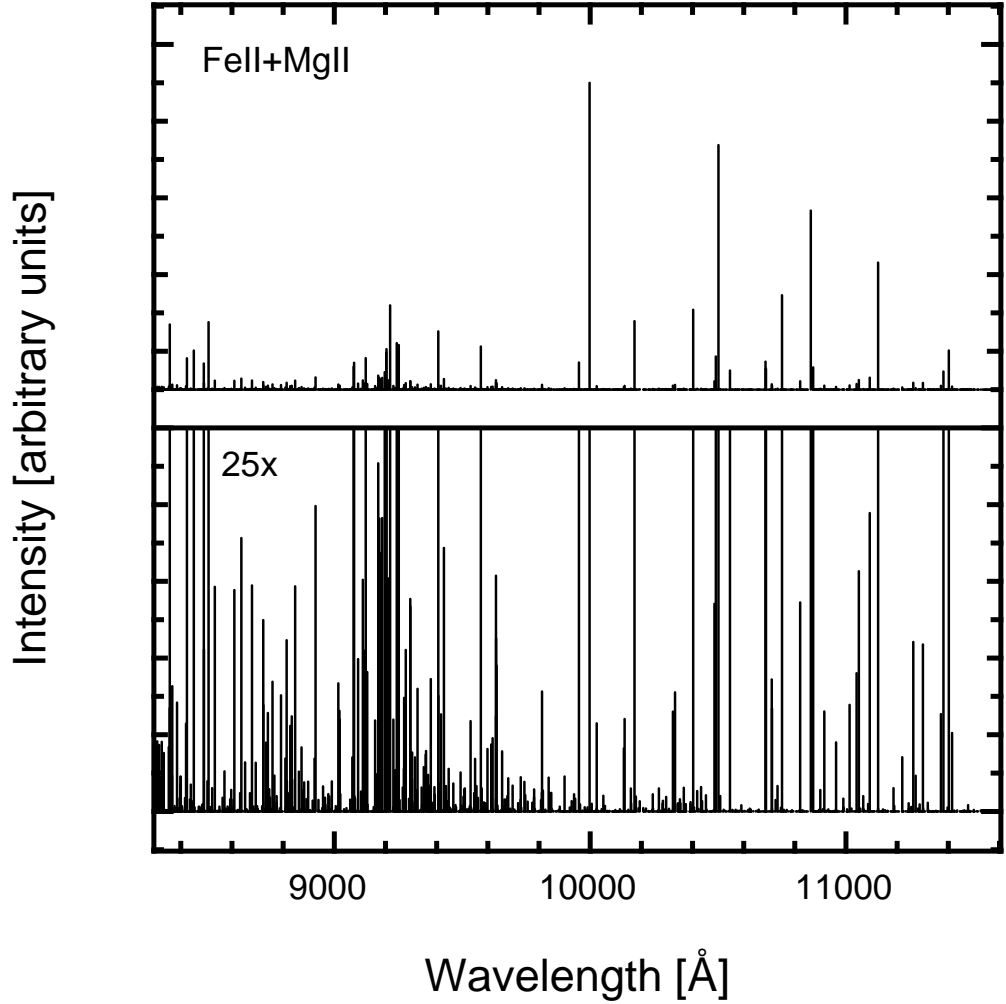}
\caption{Top: Semi-empirical template derived from the best fitted model and the deconvolution procedure applied to the observed spectrum of I\,Zw\,1. Bottom: a zoomed view of the template. Intensities are plotted in arbitrary units.}\label{fig:template}
\end{figure}

The total semi-empirical template (corresponding to $\boldsymbol{\ell}$, according to 
our notation) in the region of interest (0.83-1.16 $\mu$m) is shown in Figure \ref{fig:template}, on top. The bottom panel shows an expanded view ($\times$25) of the template, in order to highlight the contribution of the less intense lines. 

\begin{figure}
\epsscale{.80}
\plotone{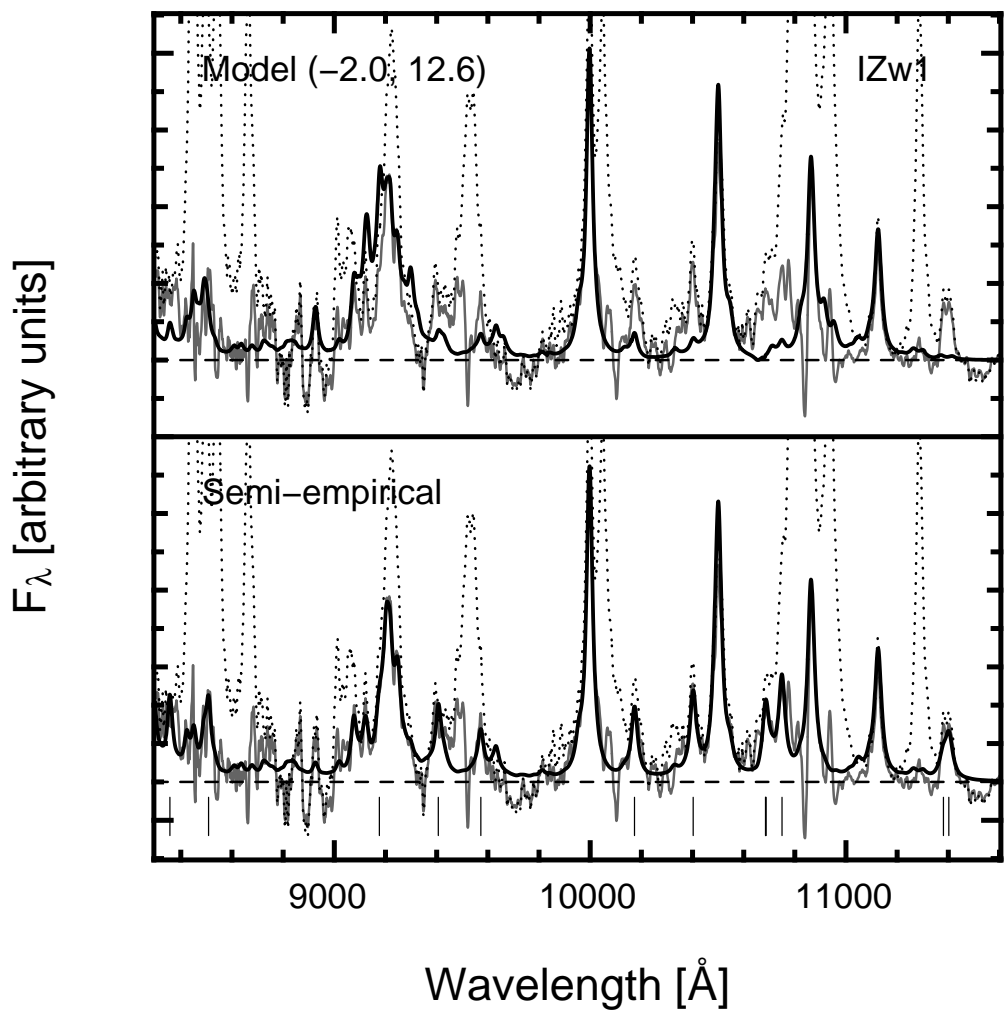}
\caption{Top: Comparison of the observed Fe\,{\sc ii}+Mg\,{\sc ii} spectrum with the one derived from the best model (-2.0,12.9). Bottom: again the observed Fe\,{\sc ii}+Mg\,{\sc ii} spectrum, and the semi-empirical spectrum, computed through the template derived in this work. Dotted black line: total emission line spectrum of I\,Zw\,1; solid grey and black lines: observed and model/semi-empirical Fe\,{\sc ii}+Mg\,{\sc ii} spectra, respectively. Dashed lines indicate the zero level intensity. Line intensities which varied significantly with respect to the model are also marked in the lower plot.}.\label{fig:obs_dec}
\end{figure}

Figure \ref{fig:obs_dec} shows the resulting semi-empirical spectrum 
(bottom), as well as the one derived from best model (top), 
for comparison, superposed on the observed Fe\,{\sc ii}+Mg\,{\sc ii} 
spectrum of I\,Zw\,1. The semi-empirical template reduces by $\sim$31\% the 
RMS of the observed spectrum with respect to the best model. 
The bottom figure also indicates the lines whose estimated intensities 
varied more significantly with respect to those given by the model. 
Notice that negative intensities obtained from the deconvolution 
process have been zeroed in the computation of the estimated spectrum, 
but are shown in table \ref{tab:dec} for completion. Such negative 
line intensities are consistent with zero within 2$\sigma$ of the 
spectral residuals. 

\cite{LEA08} [see their Section 5.4.1.] have
compared the predicted Fe\,{\sc ii} model of \citet{SP03} to
observations and found a discrepancy in the the Fe\,{\sc ii}\,1.0491+1.0502~$\mu$m
and Fe\,{\sc ii}\,1.0174\,$\mu$m emission lines, in the sense that the former were
overpredicted whereas the latter were underpredicted by theory, and
this by a similar factor of $\sim$2. Note that they used the predictions
of model A of \citet{SP03} with log\,$U_{\rm ion}$=-2 and
log\,$n_{\rm H}$=9.6. Our best matching model has the same ionization parameter
but a density that is three orders of magnitude larger (log\,$n_{\rm H}$=12.6).
As can be seen in Table~\ref{tab:el} and Figures~\ref{pfit1} and~\ref{pfit2}, 
our best model leads to a better fit of the most conspicuous lines including
Fe\,{\sc ii}\,1.0491+1.0502~$\mu$m. Indeed, the discrepancy is removed
in this pair of emission lines. Fe\,{\sc ii}\,1.0174\,$\mu$m, on the other hand, continues 
to be underpredicted even in the best model. Actually, this was one of the iron features that needed a fine tuning:
its strength increased in the semi-empirical template by a factor of
almost 3 (see Figure~\ref{fig:izw1_final}, when a comparison between the model 
and the semi-empirical template is done). Very likely, atomic parameters
for that line would need to be reviewed in future modeling.

\subsection{Application to Ark\,564}

Ark\,564 is another well-known NLS1 with strong Fe\,{\sc ii} 
emission, suitable to test our semi-empirical template. The spectrum of 
this AGN was observed and reduced in a similar way to that of of I\,Zw\,1.
Details of the extraction, reduction, wavelength and flux calibration
are in \cite{RA04}. 
As previously, we fitted and subtracted the continuum emission through a spline 
function. For modeling the line profiles, we adopted Pa$\alpha$ as representative
of the emission line profile to be convolved with the Fe\,{\sc ii}+Mg\,{\sc ii} 
intensity vector 
derived from I\,Zw\,1, and also as a template for other permitted emission 
lines. Forbidden lines were assumed to have Gaussian\footnote{Ark\,564 
shows sharper forbidden emission features in its spectrum, so we decided 
to use a Gaussian instead of a Lorentzian to model them.} profiles with 
widths of 500-550 km~s$^{-1}$. A quick fit over all the theoretical models 
also indicates a very high density and low/moderate ionization parameter 
for the Fe\,{\sc ii} emitting region, as shown in Table \ref{tab:akn564} with the 
estimated RMS over the 8300-11600 \AA~ region. The sharpness of the 
emission lines makes small uncertainties in emission line positioning and 
scaling to give rise to high amplitude residuals in the fit, what 
explains the larger observed residual RMS. Neverthless, the general trend 
is similar to that observed for I\,Zw\,1, favoring similar 
physical conditions for the Fe\,{\sc ii} emitting region. 

\begin{figure}
\epsscale{.80}
\plotone{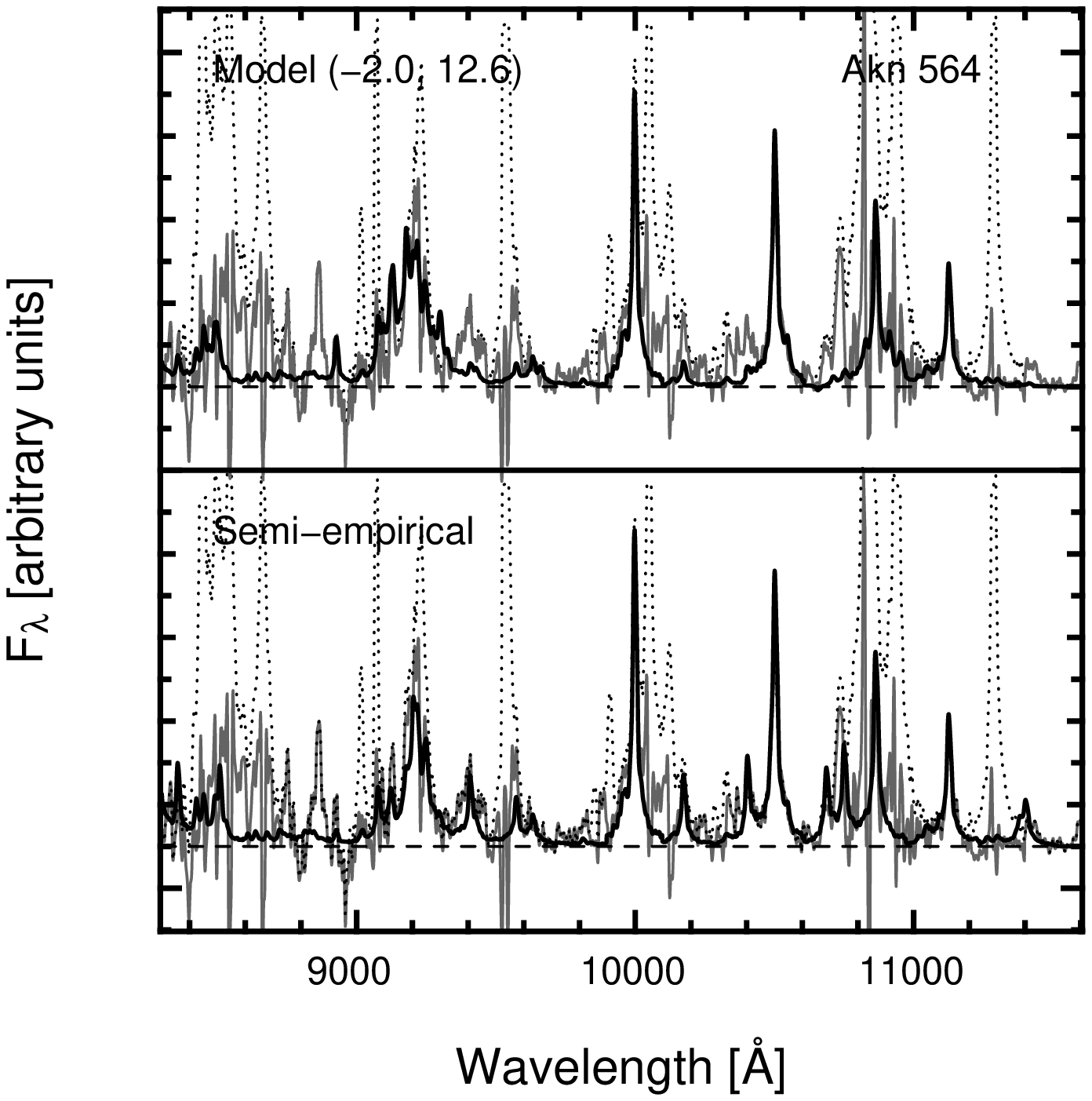}
\caption{Top: Comparison of the observed Fe\,{\sc ii}+Mg\,{\sc ii} spectrum of Ark\,564 with the one derived 
from the best model (-2.0,12.9). Bottom: again the observed Fe\,{\sc ii}+Mg\,{\sc ii} spectrum, and
the semi-empirical spectrum, computed through the template derived from I\,Zw\,1. Dotted black line: total 
emission line spectrum of Ark\,564; solid grey and black lines: observed and model/semi-empirical FeII+MgII spectra, respectively. Dashed lines: zero intensity level.}.\label{fig:akn564}
\end{figure}

Figure \ref{fig:akn564} is similar to figure \ref{fig:obs_dec}, but 
now showing the suitability of our derived semi-empirical template to the 
observed Fe\,{\sc ii}+Mg\,{\sc ii} emission of Ark\,564. A smaller RMS in the 
residuals with respect to those obtained from the theoretical models confirms 
the usefulness of the semi-empirical template to represent the Fe\,{\sc ii} 
and Mg\,{\sc ii} emission in AGNs. In general, lines that needed a
fine tuning in I\,Zw\,1, like Fe\,{\sc ii}\,10174 \AA, properly
reproduce the observations. Note, however, that the bump 
observed at 11400~\AA~in I\,Zw\,1 and that we deliberately introduced in the 
template seems to be unusually large with 
respect to the one in Ark\,564. Clearly, the transitions leading to
these particular set of lines should strongly depend on local physical 
parameters that vary from object to object.  

\section{Discussion}

The analysis carried out in the previous sections confirms 
several pieces of evidence already suggested by other works,
which are: {\it (i)} Ly$\alpha$ fluorescence is indeed a process
that should be taken into account in any systematic study of the
Fe\,{\sc ii} emission in AGNs as it produces
a considerable amount of emission lines that otherwise would be
absent. This is particularly evident for the 9200~\AA\ feature,
composed of numerous Fe\,{\sc ii} multiples as well as some 
contribution from Mg\,{\sc ii} lines.  {\it (ii)} Similarly to the optical 
region, the NIR Fe\,{\sc ii} emission also produces a subtle pseudo-continuum, 
particularly in the region between 8600 and 10000~\AA\ 
(see Figure~\ref{fig:template}). Without a proper modeling and subtraction 
of this emission, fluxes of other BLR and NLR features can be 
severely overestimated. {\it (3)} Unlike in the optical region,
individual Fe\,{\sc ii} emission lines can be isolated 
in the NIR (i.e., the lines at 10501~\AA, 10862~\AA\ and 11126~\AA).
This is particularly useful to characterize, for instance, the
line profiles and emission line fluxes, in an already complex
emission region. This individual Fe\,{\sc ii} line charaterization
can be possible because other Fe\,{\sc ii} lines very close 
in wavelength to the above three are at least 25$\times$ weaker. 

\begin{figure}
\epsscale{.80}
\plotone{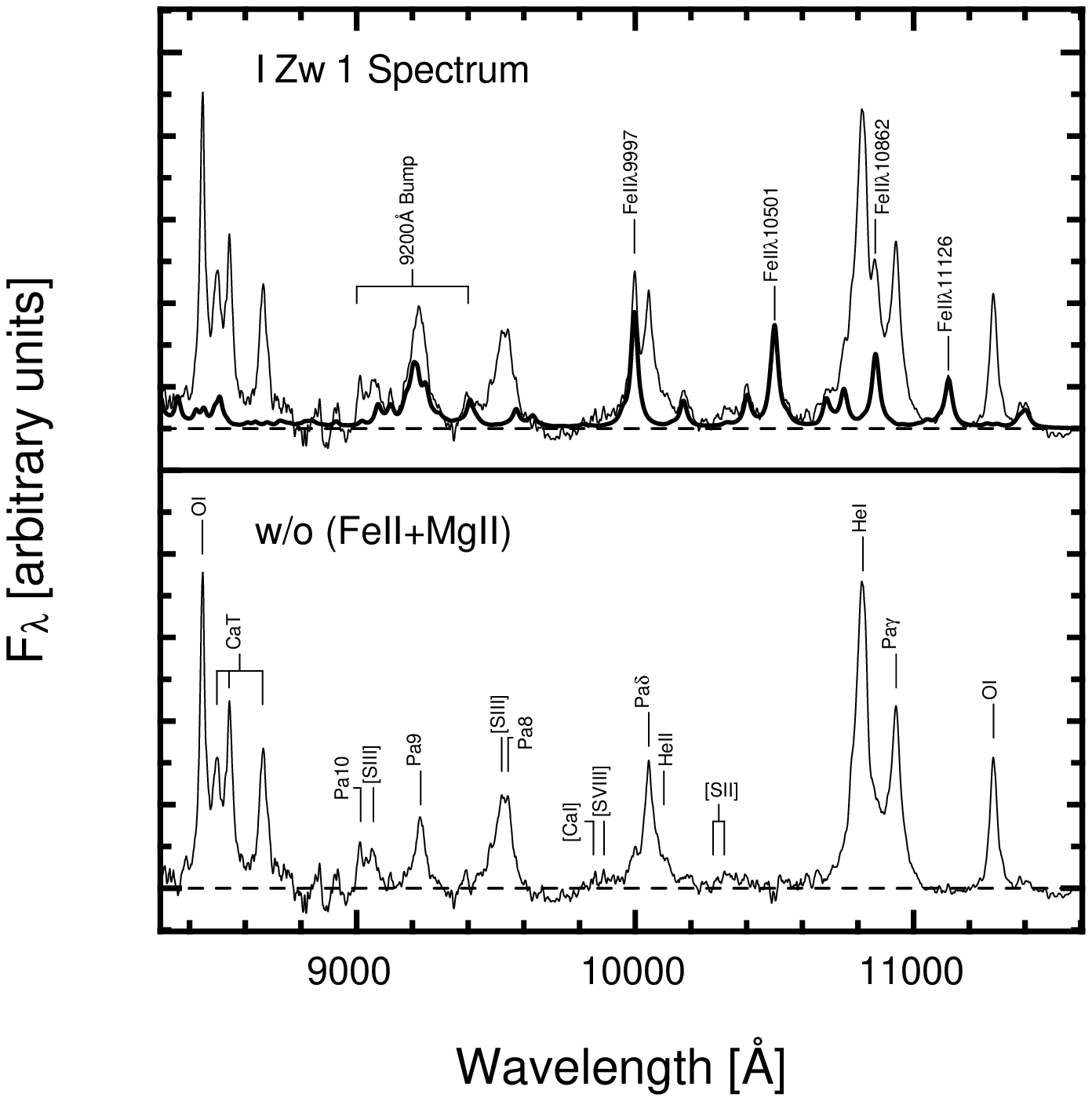}
\caption{Top: Continuum-subtracted spectrum of I\,Zw\,1, with the spectrum of Fe\,{\sc ii}+Mg\,{\sc ii} calculated from the semi-empirical template (in bold) superposed, highlighting its main NIR emission lines. Bottom: Spectrum of I\,Zw\,1 without this contribution. All the lines (except Pa$\alpha$ and those of the template) used in the fit are marked.}\label{fig:izw1_final}
\end{figure}

The above thoughts can better be visualized in Figure~\ref{fig:izw1_final}, 
which shows the spectrum of I\,Zw\,1 with (top) and without (bottom) 
the Fe\,{\sc ii}+Mg\,{\sc ii} contribution. The spectrum ``clean'' of iron 
and magnesium emission highlights the remaining emission lines, properly identified in 
the figure. Small spikes, coincident with the position of the strongest Fe\,{\sc ii} 
lines can still be seen as, for example, in the blue part of Pa$\delta$, 
between He\,{\sc i} and Pa$\gamma$ and at the position of the 
Fe\,{\sc ii}\,$\lambda$11126 line. We interpret these small residuals 
as a due to emission from the NLR, as suggested by \cite{VCJV04}. It might indicate that a more complex modeling 
of the convolving profile might be required, for example, by including
a possible contribution from the NLR. However, because we are
primarily interested in the construction of a semi-empirical template
for the Fe\,{\sc ii}+Mg\,{\sc ii} emission which originates at the BLR,
these small residuals are out of concern. 

\begin{figure}
\epsscale{.80}
\plotone{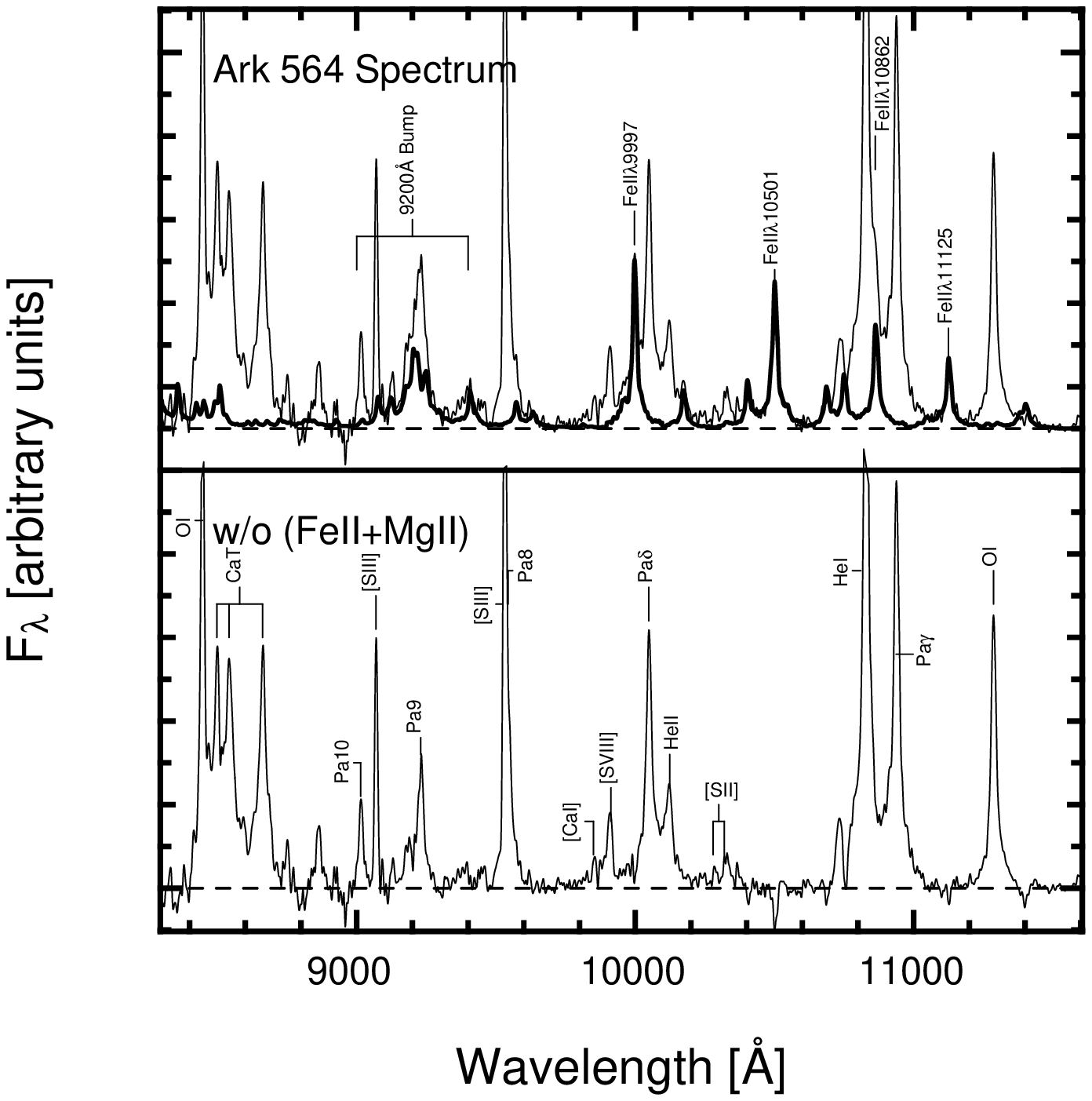}
\caption{Top: Continuum-subtracted spectrum of Ark\,564, with the spectrum of Fe\,{\sc ii}+Mg\,{\sc ii} calculated from the semi-empirical template (in bold) superposed, highlighting its main NIR emission lines. Bottom: Spectrum of Ark\,564 without this contribution. All the lines (except Pa$\alpha$ and those of the template) used in the fit are marked.}\label{fig:ark564_final}
\end{figure}

Figure \ref{fig:ark564_final} shows the 
same as figure \ref{fig:izw1_final} but for Ark\,564. It can be seen
that the spectrum at the bottom is nicely clean of Fe\,{\sc ii} and 
Mg\,{\sc ii}, as evidenced by the small residuals left in the region 
between 10200-10600~\AA. It exemplifies the use of our NIR I\,Zw\,1 
template to remove that emission in other AGNs.

At this point we call the attention to an apparent absorption
feature blueward of He\,{\sc i}\,$\lambda$10818 that appeared 
after the subtraction of the Fe\,{\sc ii}+Mg\,{\sc ii} 
semi-empirical template in Ark\,564. It is possible
that this feature is artificial and due to a bad subtraction of the 
Fe\,{\sc ii} blends with peak at 10750\,\AA. Yet another possibility
is that it could be a real feature, similar to the one reported by \cite{LDB11} in 
the quasar \objectname[FBQS J1151+3822]{FBQSJ1151+3822}, which they 
attribute to a broad absorption line (BAL) system
in that source. \cite{LDB11} discussed the prospects of finding 
other He\,{\sc i}\,$\lambda$10830 BALQSOs on six additional objects
and pointed out that several well-known, bright 
low-redshift BALQSOs have no He\,{\sc i}\,$\lambda$10830 absorption,
a fact that can place upper limits on the column densities in those
objects. Observations with higher spectral resolution are needed to confirm if 
the absorption in Ark\,564 is indeed real.
Although it is out of the scope of this paper the
study of such a system of absorbers in the BLR of Ark\,564 or in
other sources, our results strengthen the need of an adequate 
Fe\,{\sc ii} subtraction around He\,{\sc i}\,$\lambda$10830 to further 
constrain the inner physical properties of such AGNs.  

Note also that the peak observed around 10740~\AA~ is due 
to [Fe\,{\sc xiii}] emission. In addition, the 
11400~\AA\ bump is overestimated in the semi-empirical template,
meaning that some particular lines may need a fine tuning for
a better match. It implies also that not all NIR Fe\,{\sc ii} lines 
may scale up by the same factor, as would be expected. However,
as in the optical and UV regions, the semi-empirical template suitably reproduces most of the observed Fe\,{\sc ii}+Mg\,{\sc ii} emission features in the NIR. Clearly, testing the 
template in a large number of objects is necessary to 
verify its suitability in more general terms.

At this point is important to draw our attention to how significant is the
NIR Fe\,{\sc ii} emission in  I\,Zw\,1 compared to that of
the optical and UV. For this purpose we measured the integrated
flux in the 8300$-$11600~\AA\ region using the Fe\,{\sc ii}
semi-empirical template derived for that object. We found that
F$_{Fe\,{\sc II}}=2.79\pm0.20\times10^{-13}$~erg\,cm$^{-2}$\,s$^{-1}$.
Note that the Mg\,{\sc ii} was not taken into account in the
computed value. \citet{tsu06} measured the integrated flux
of Fe\,{\sc ii} for I\,Zw\,1 using HST and ground-based observations
of this galaxy in five different wavelength bands:
$U$1 [2200-2660~\AA], $U$2 [2660-3000~\AA], $U$3 [3000-3500~\AA],
$O$1 [4400-4700~\AA] and $O$2 [5100-5600~\AA]. The values
they found, relative to H$\beta$, are shown in
Table~\ref{tab:fe2flux}.

It can be seen that nearly half the amount of H$\beta$ flux
in I\,Zw\,1 is emitted by Fe\,{\sc ii} in the NIR. Moreover,
the integrated NIR iron emission carries a flux that is
equivalent to $\sim$10\% of the Fe\,{\sc ii} emission in
the optical region (sum of the fluxes in the $O$1 and
$O$2 intervals of Table~\ref{tab:fe2flux}),
to $\sim$10\% of the Fe\,{\sc ii} in the near-UV region
(3000-3500~\AA) and to $\sim$3\% of the Fe\,{\sc ii} emission
in the UV (2200-3000~\AA).

At first sight the above numbers may indicate
that the role of Ly$\alpha$ pumping, which is behind most
of the Fe\,{\sc ii} NIR emission, is negligible.
However, we should take into account that after the cascading
transitions that result in the NIR emission the z~$^4$D and z~$^4$F levels
are populated. These latter are responsible for
part of the transitions leading to the Fe\,{\sc ii} emission
in the $O$1 and $O$2 optical regions.
Therefore, a 10\% in flux means that up to 20\% of the
optical Fe\,{\sc ii} photons can be attributed to Ly$\alpha$ fluorescence.

\section{Conclusions}

We have compared theoretical models for the Fe\,{\sc ii}+Mg\,{\sc ii} 
NIR emission in active galaxies, which have the same turbulent velocities 
in the medium, but differ in physical conditions such as density and degree 
of ionization of the emitting region. For that purpose, we chose to model the NLS1 
galaxy I\,Zw\,1, which has traditionally provided the template for the Fe\,{\sc ii} 
emission in the optical and UV. 

The best match among all models was obtained by comparing the results of 
a multi-parametric fit comprising the main emission line features to the observed 
spectrum in the region 0.83-1.16 $\mu$m. Low/moderate ionization parameters and 
high gas densities of 10$^{12.6}$~cm$^{-3}$) are favored, 
as they reduce the residuals of the fitted spectrum. However, since some 
Fe\,{\sc ii} lines are clearly underestimated even by the best models, we 
decided to derive a semi-empirical template to improve the fitting, by adjusting 
the intensities of the most prominent Fe\,{\sc ii} and Mg\,{\sc ii} lines, taken 
from the best fitted model. 

The newly derived template reduces the fit residuals by about 31\% with 
respect to what is obtained using the best theoretical model. We performed a quick check on the spectrum of another NLS1 with conspicuous and very narrow emission lines, Ark\,564. Despite some small differences, this test corroborated the reliability of this new 
semi-empirical template to reproduce AGN Fe\,{\sc ii} and Mg\,{\sc ii} emission lines in the NIR.

We also highlight that the Fe\,{\sc ii} bump around 11400 \AA\ which we introduced to match the observed spectrum of I\,Zw\,1 does not reproduce well the same observed feature of Ark\,564. This lead us to the conclusion that this emission is abnormal in I\,Zw\,1, given that none of the models too could predict such a large emission. Also, I\,Zw\,1 might contain a narrow contribution to the Fe\,{\sc ii} spectrum, as suggested in an optical study of \cite{VCJV04}.  Further tests will be futurely carried out on a larger sample, in order to fine tune our derived template.

\acknowledgments AGR acknowledges Instituto Nacional de Ci\^encia e Tecnologia de Astrof\'{i}sica (INCT-A) for funding support under process CPNq 573648/2008-5. ARA acknowledges CNPq for partial support to this research through grant 308877/2009-8. AKP would like to acknowledge partial support from the U.S. National Science Foundation, and Sultana Nahar for the Iron Project data. We thank to an anonymous Referee for useful corrections suggested to the manuscript.

%
\begin{deluxetable}{lrcl}
\tabletypesize{\scriptsize}
\tablecaption{Template lines that were added to the Fe\,{\sc ii}+Mg\,{\sc ii} fitting process.\label{tab:prof}}
\tablewidth{0pt}
\tablehead{
\colhead{Line} & \colhead{$\lambda_{rest}$ (\AA)} & \colhead{$\lambda_{meas}$ (\AA)} & \colhead{Profile} }
\startdata  
Pa$\alpha$  &  18750          & 18750 &	    itself    \\
Pa$\gamma$  &  10937          & 10937 &     Pa$\alpha$ \\
Pa$\delta$  &  10049          & 10049 &     Pa$\alpha$ \\
Pa8        &   9544          &  9544 &     Pa$\alpha$ \\
Pa9        &   9230          &  9230 &     Pa$\alpha$ \\
Pa10       &   9014          &  9014 &     Pa$\alpha$ \\
Ca\,{\sc ii}        &   8498          &  8498 &     Pa$\alpha$ \\
Ca\,{\sc ii}        &   8542          &  8542 &     Pa$\alpha$ \\
Ca\,{\sc ii}        &   8662          &  8662 &     Pa$\alpha$ \\
O\,{\sc i}          &   8446          &  8446 &     OI$\lambda$11287 \\
O\,{\sc i}          &  11287          & 11287 &     OI$\lambda$11287 \\
He\,{\sc i}         &  10830          & 10818 &     Pa$\alpha$ \\
He\,{\sc ii}        &  10124          & 10103 &     Pa$\alpha$ \\
$[$CaI$]$   &   9850          & 9857  &  Lorentzian   \\
$[$S\,{\sc ii}$]$   &  10280          & 10280 &  Lorentzian   \\
$[$S\,{\sc ii}$]$   &  10320          & 10320 &  Lorentzian   \\
$[$S\,{\sc iii}$]$\tablenotemark{a}   &   9069          & 9060  &  Lorentzian   \\
$[$S\,{\sc iii}$]$\tablenotemark{a}   &   9532          & 9521  &  Lorentzian   \\
$[$S\,{\sc viii}$]$ &   9913          & 9888  &  Lorentzian   \\
\enddata
\tablenotetext{a}{~ Line peak intensity ratio fixed in 2.4.}
\end{deluxetable}

\begin{deluxetable}{lcccccccccccc}
\rotate
\tabletypesize{\scriptsize}
\tablecaption{Fluxes of I\,Zw\,1 emission lines obtained from the fitting of Fe\,{\sc ii}+Mg\,{\sc ii} (in units of 10$^{-14}$ erg s$^{-1}$ cm$^{-2}$).\label{tab:el}}
\tablewidth{0pt}
\tablehead{
  &  \multicolumn{4}{c}{log $U_{\rm ion}$=-3}   & \multicolumn{4}{c}{log $U_{\rm ion}$=-2}  & \multicolumn{4}{c}{log $U_{\rm ion}$=-1.3} \\
\colhead{log $n_{\rm H}$}  & \colhead{9.6} &  \colhead{10.6}  &  \colhead{11.6}  & \colhead{12.6}   & \colhead{9.6} &  \colhead{10.6}  &  
		\colhead{11.6}  & \colhead{12.6}   & \colhead{9.6} &  \colhead{10.6}  &  \colhead{11.6}  & \colhead{12.6}} 
\startdata  
Pa$\gamma$ (10937\AA)   & 9.18 & 9.16 & 9.01 & 8.64 & 8.81 & 8.96 & 8.64 & 8.64 & 8.82 & 8.64 & 8.64 & 8.64\\
Pa$\delta$ (10049\AA)   & 4.41 & 4.41 & 4.41 & 4.41 & 4.41 & 4.41 & 4.41 & 4.41 & 4.41 & 4.41 & 4.41 & 4.41\\
Pa8 (9544\AA)   	& 2.78 & 2.78 & 2.78 & 2.78 & 2.78 & 2.78 & 2.78 & 2.78 & 2.78 & 2.78 & 2.78 & 2.78\\
Pa9 (9230\AA)   	& 2.22 & 2.22 & 2.22 & 2.22 & 2.22 & 2.22 & 2.22 & 2.22 & 2.22 & 2.22 & 2.17 & 2.17\\ 
Pa10 (9014\AA)  	& 1.43 & 1.43 & 1.39 & 1.26 & 1.38 & 1.40 & 1.26 & 1.24 & 1.38 & 1.32 & 1.17 & 1.12\\ 
He\,{\sc i} (10830\AA)       	      & 18.9 & 18.9 & 18.7 & 17.8 & 18.6 & 18.8 & 18.0 & 17.3 & 18.7 & 18.4 & 17.3 & 17.4\\ 
He\,{\sc ii} (10124\AA)      	      & 1.73 & 1.73 & 1.75 & 1.59 & 1.61 & 1.73 & 1.63 & 1.39 & 1.63 & 1.67 & 1.49 & 1.40\\ 
O\,{\sc i}  (8446\AA)        	      & 7.21 & 7.21 & 7.15 & 7.03 & 7.14 & 7.17 & 7.04 & 7.30 & 7.14 & 7.15 & 7.28 & 7.30\\
O\,{\sc i}  (11287\AA)       	      & 4.62 & 4.63 & 4.62 & 4.52 & 4.60 & 4.62 & 4.55 & 4.47 & 4.62 & 4.58 & 4.50 & 4.44\\
Ca\,{\sc ii} (8498\AA)        	      & 3.63 & 3.62 & 3.54 & 3.06 & 3.57 & 3.57 & 3.21 & 3.09 & 3.56 & 3.45 & 3.19 & 3.00\\
Ca\,{\sc ii} (8542\AA)        	      & 5.93 & 5.93 & 5.92 & 5.86 & 5.92 & 5.92 & 5.88 & 5.79 & 5.92 & 5.90 & 5.83 & 5.75\\
Ca\,{\sc ii} (8662\AA)        	      & 5.19 & 5.19 & 5.18 & 5.10 & 5.17 & 5.18 & 5.12 & 4.95 & 5.17 & 5.16 & 5.04 & 4.83\\
$[$Ca\,{\sc i}$]$ (9850\AA)           & 0.25 & 0.25 & 0.25 & 0.24 & 0.24 & 0.25 & 0.24 & 0.22 & 0.24 & 0.24 & 0.23 & 0.23\\ 
$[$S\,{\sc ii}$]$ (10280+10320\AA)    & 1.00 & 1.00 & 0.99 & 0.87 & 0.98 & 1.00 & 0.91 & 0.79 & 0.99 & 0.96 & 0.84 & 0.84\\
$[$S\,{\sc iii}$]$ (9530+9068\AA)     & 3.03 & 3.03 & 2.99 & 2.85 & 2.96 & 2.99 & 2.86 & 2.89 & 2.96 & 2.93 & 2.88 & 2.83\\
$[$S\,{\sc viii}$]$ (9913\AA)         & 0.51 & 0.51 & 0.50 & 0.43 & 0.49 & 0.50 & 0.45 & 0.35 & 0.49 & 0.48 & 0.39 & 0.35\\
Fe\,{\sc ii} (9997\AA)                & 0.34 & 0.34 & 0.45 & 1.99 & 0.69 & 0.45 & 1.49 & 3.78 & 0.68 & 0.84 & 2.77 & 3.73 \\ 
Fe\,{\sc ii} (10501\AA)               & 0.35 & 0.36 & 0.50 & 2.06 & 0.74 & 0.49 & 1.68 & 3.16 & 0.74 & 0.99 & 2.37 & 1.94 \\ 
Fe\,{\sc ii} (10863\AA)               & 0.28 & 0.31 & 0.43 & 1.69 & 0.65 & 0.42 & 1.43 & 2.41 & 0.67 & 0.85 & 2.33 & 1.86 \\
Fe\,{\sc ii} (11126\AA)               & 0.14 & 0.16 & 0.25 & 1.11 & 0.33 & 0.23 & 0.89 & 1.70 & 0.34 & 0.49 & 1.51 & 1.37 \\ 
Fe\,{\sc ii} (9000-9400\AA)\tablenotemark{a}           & 6.37 & 6.38 & 6.81 & 8.93 & 7.14 & 6.70 & 8.53 & 7.80 & 7.12 & 7.23 & 7.82 & 8.44 \\ 
Mg\,{\sc ii} (8300-11600\AA)\tablenotemark{a}      & 0.93 & 0.97 & 1.32 & 2.47 & 1.74 & 1.51 & 2.46 & 2.75 & 1.75 & 3.25 & 3.51 & 3.77 \\
\enddata
\tablenotetext{a}{Integrated fluxes of all lines in the interval.}
\end{deluxetable}

\begin{deluxetable}{rcclcrccl}
\tabletypesize{\scriptsize}
\tablecaption{Relative intensities of Fe\,{\sc ii} and Mg\,{\sc ii} lines in the $\boldsymbol{\ell}^*$ vector as
given by the original best model (-2.0, 12.6) and by the one computed from the observed spectrum. Values have been normalized with respect to the intensity of the 9997 \AA~ line.\label{tab:dec}}
\tablewidth{0pt}
\tablehead{
\colhead{$\lambda$ (\AA)} & \colhead{$\boldsymbol{\ell}^*_{model}$} & \colhead{$\boldsymbol{\ell}^*_{deconv}$} & \colhead{Ion} & \colhead{~~~} & \colhead{$\lambda$ (\AA)} & \colhead{$\boldsymbol{\ell}^*_{model}$} & \colhead{$\boldsymbol{\ell}^*_{deconv}$} & \colhead{Ion}}
\startdata
 8213.99 & 0.055 & 0.032 & Mg\,{\sc ii} & &   9204.05 & 0.093 & 0.132 & Fe\,{\sc ii} \\
 8228.93 & 0.057 & 0.043 & Fe\,{\sc ii} & &   9218.25 & 0.304 & 0.274 & Mg\,{\sc ii} \\
 8234.64 & 0.111 & 0.049 & Mg\,{\sc ii} & &   9244.26 & 0.186 & 0.151 & Mg\,{\sc ii} \\
 8287.85 & 0.103 & 0.350 & Fe\,{\sc ii} & &   9251.72 & 0.080 & 0.146 & Fe\,{\sc ii}   \\ 
 8357.18 & 0.062 & 0.212 & Fe\,{\sc ii} & &   9272.16 & 0.062 & 0.006 & Fe\,{\sc ii}   \\ 
 8423.87 & 0.076 & 0.102 & Fe\,{\sc ii} & &   9296.85 & 0.046 & 0.028 & Fe\,{\sc ii}   \\ 
 8450.99 & 0.160 & 0.127 & Fe\,{\sc ii} & &   9297.23 & 0.102 & 0.027 & Fe\,{\sc ii}   \\
 8469.22 & 0.066 & -0.050 & Fe\,{\sc ii}& &   9303.59 & 0.071 & 0.004 & Fe\,{\sc ii}   \\  
 8490.05 & 0.124 & 0.085 & Fe\,{\sc ii} & &   9326.93 & 0.048 & -0.089 & Fe\,{\sc ii}  \\ 
 8499.56 & 0.091 & -0.034 & Fe\,{\sc ii}& &   9406.67 & 0.041 & 0.190 & Fe\,{\sc ii}   \\
 8508.61 & 0.051 & 0.220 & Fe\,{\sc ii} & &   9572.62 & 0.061 & 0.140 & Fe\,{\sc ii}   \\  
 8926.64 & 0.158 & 0.040 & Fe\,{\sc ii} & &   9661.15 & 0.041 & -0.107 & Fe\,{\sc ii}  \\  
 9075.50 & 0.102 & 0.074 & Fe\,{\sc ii} & &   9956.25 & 0.103 & 0.088 & Fe\,{\sc ii}   \\ 
 9077.40 & 0.080 & 0.088 & Fe\,{\sc ii} & &  10173.51 & 0.081 & 0.223 & Fe\,{\sc ii}  \\
 9095.07 & 0.087 & -0.007 & Fe\,{\sc ii}& &  10402.83 & 0.042 & 0.260 & Fe\,{\sc ii}  \\
 9122.94 & 0.202 & 0.103 & Fe\,{\sc ii} & &  10546.38 & 0.096 & 0.062 & Fe\,{\sc ii}  \\
 9132.36 & 0.168 & -0.007 & Fe\,{\sc ii}& &  10685.17 & $\leq$0.001 & 0.091 & Fe\,{\sc ii}\tablenotemark{a} \\
 9155.77 & 0.071 & -0.022 & Fe\,{\sc ii}& &  10686.80 & $\leq$0.001 & 0.069 & Fe\,{\sc ii}\tablenotemark{a}  \\
 9171.62 & 0.040 & 0.045 & Fe\,{\sc ii} & &  10686.92 & $\leq$0.001 & 0.067 & Fe\,{\sc ii}\tablenotemark{a} \\  
 9175.87 & 0.169 & 0.038 & Fe\,{\sc ii} & &  10749.72 & 0.042 & 0.308 & Fe\,{\sc ii}  \\			
 9178.09 & 0.092 & 0.034 & Fe\,{\sc ii} & &  10826.50 & 0.060 & -0.159 & Fe\,{\sc ii} \\
 9179.47 & 0.103 & 0.032 & Fe\,{\sc ii} & &  10914.24 & 0.110 & -0.014 & Mg\,{\sc ii} \\
 9187.16 & 0.086 & 0.038 & Fe\,{\sc ii} & &  10951.78 & 0.078 & -0.021 & Mg\,{\sc ii} \\ 
 9196.90 & 0.055 & 0.057 & Fe\,{\sc ii} & &  11381.00 & $\leq$0.001 & 0.059 & Fe\,{\sc ii}\tablenotemark{a} \\
 9203.12 & 0.075 & 0.120 & Fe\,{\sc ii} & &  11402.00 & $\leq$0.001 & 0.127 & Fe\,{\sc ii}\tablenotemark{a} \\
\enddata
\tablenotetext{a}{~Lines included by visual inspection of the residual spectrum, after applying the threshold criterium for line selection.}
\end{deluxetable}

\begin{deluxetable}{lccc}
\tabletypesize{\scriptsize}
\tablecaption{Residual RMS (in 10$^{-15}$ erg s$^{-1}$ cm$^{-2}$ \AA$^{-1}$) obtained from the fit of Ark\,564, using the models and the template derived from I\,Zw\,1.\label{tab:akn564}}
\tablewidth{0pt}
\tablehead{
& \multicolumn{3}{c}{log $U_{\rm ion}$}  \\
 & -3.0 & -2.0 & -1.3}
 \startdata
log $n_{\rm H}$=9.6  & 0.466 & 0.447 & 0.447\\
log $n_{\rm H}$=10.6 & 0.459 & 0.462 & 0.436\\
log $n_{\rm H}$=11.6 & 0.455 & 0.421 & 0.396\\
log $n_{\rm H}$=12.6 & 0.409 & 0.392 & 0.392\\
& & & \\
Semi-empirical & & 0.377 & \\
\enddata
\end{deluxetable}

\begin{deluxetable}{lcc}
\tabletypesize{\scriptsize}
\tablecaption{Integrated emission line fluxes for I\,Zw\,1 in the UV, optical and near-infrared regions.\label{tab:fe2flux}}
\tablewidth{0pt}
\tablehead{
\colhead{Spectral Region\tablenotemark{a}} & \colhead{Integrated Flux Ratio\tablenotemark{b}} & \colhead{Reference}}
\startdata
$U$1 (2200-2660) & 8.06$\pm$0.39 & \cite{tsu06}\\
$U$2 (2660-3000) & 5.45$\pm$0.27 & \cite{tsu06}\\
$U$3 (3000-3500) & 4.50$\pm$0.23 & \cite{tsu06}\\
$O$1 (4400-4700) & 2.65$\pm$0.18 & \cite{tsu06}\\
$O$2 (5100-5600) & 2.68$\pm$0.16 & \cite{tsu06}\\
NIR  (8300-11600) &0.46$\pm$0.02 & this work\\
\enddata
\tablenotetext{a}{$U$1, $U$2, $U$3, $O$1, $O$2 and NIR denote the integrated Fe\,{\sc ii} emission in the intervals (in \AA) shown between brackets, relative to H$\beta$ flux.}
\tablenotetext{b}{Values are relative to the integrated H$\beta$ flux of 6.08$\pm$0.24 erg\,cm$^{-2}$\,s$^{-1}$ \citep{tsu06}.}
\end{deluxetable}


\begin{figure}
\epsscale{.80}
\plotone{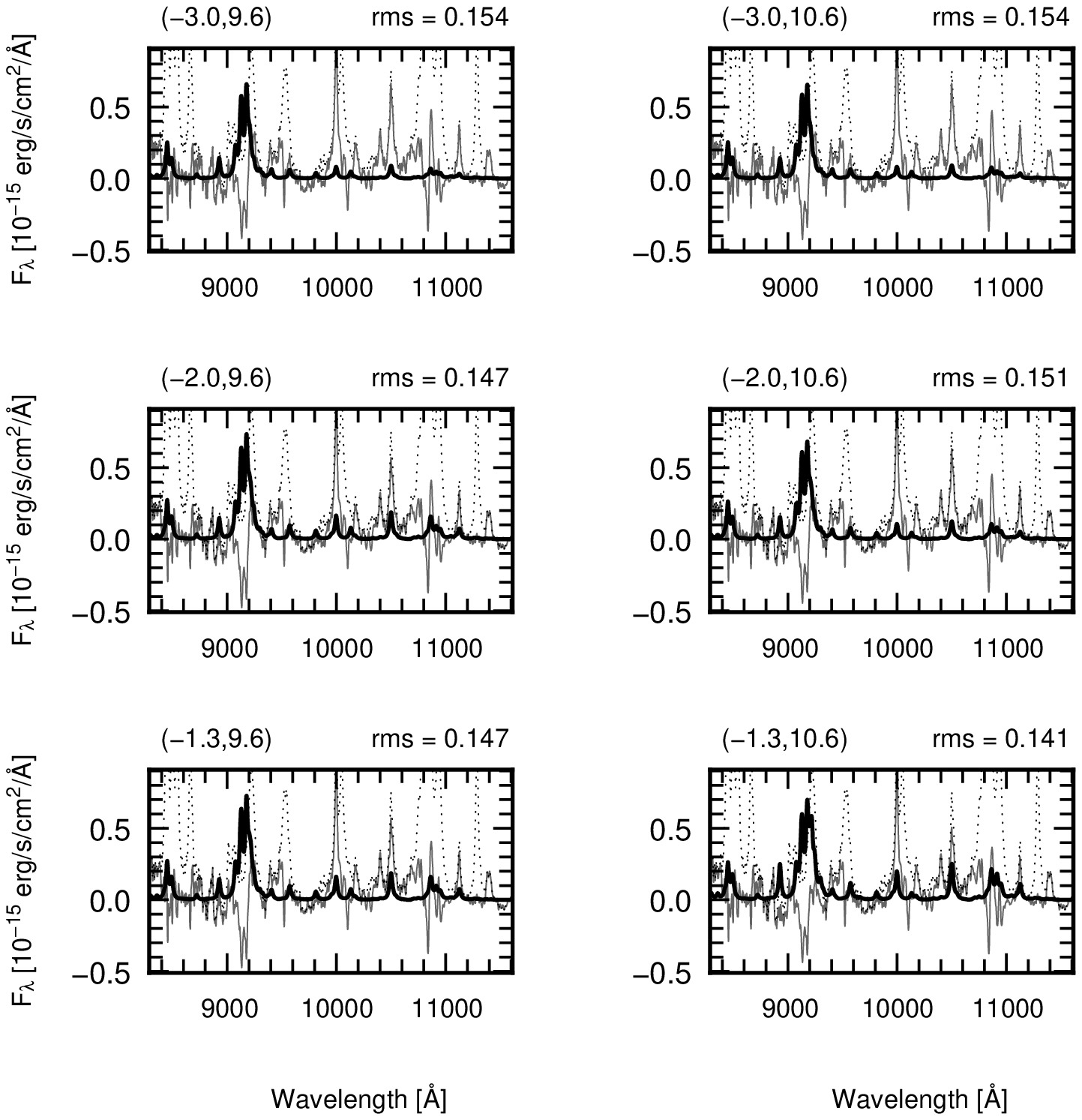}
\caption{Results of the fitting procedure for the templates of Fe\,{\sc ii}+Mg\,{\sc ii}, for 
$\log n_{\rm H}$ of 9.6 and 10.6. Dotted lines are the continuum-subtracted spectrum of I\,Zw\,1, solid-bold 
lines show the fitted (convolved) template, and the lighter solid lines are the fit residuals.}.\label{pfit1}
\end{figure}

\begin{figure}
\epsscale{.80}
\plotone{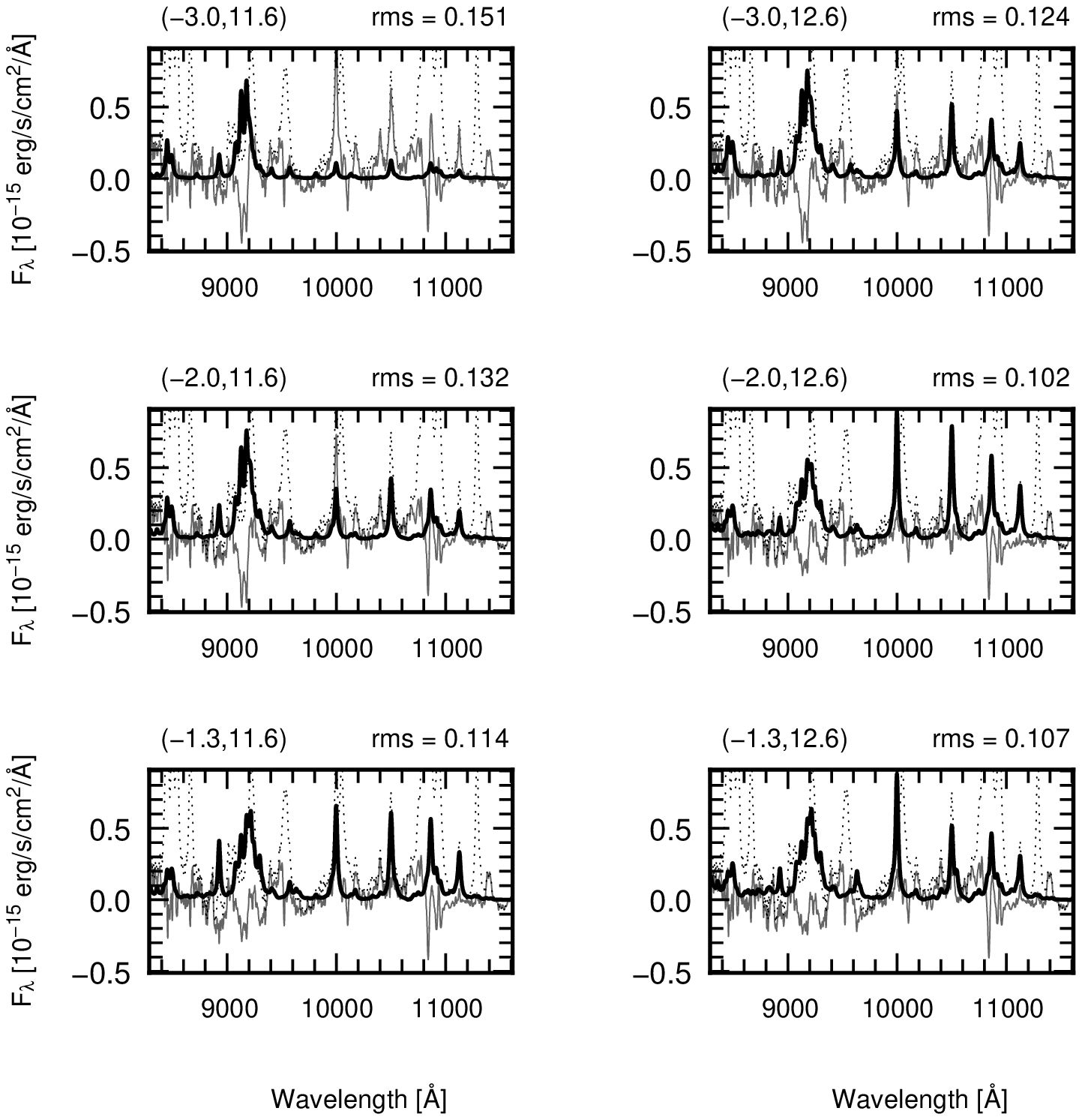}
\caption{(Cont.): results of the fitting procedure for the templates of Fe\,{\sc ii}+Mg\,{\sc ii}, for $\log n_{\rm H}$ of 11.6 and 12.6. Dotted lines are the continuum-subtracted spectrum of I\,Zw\,1, 
solid-bold lines show the fitted (convolved) template, and the lighter solid lines are the  fit residuals.}.\label{pfit2}
\end{figure}

\end{document}